\documentclass[fleqn,usenatbib]{mnras}

\usepackage{newtxtext,newtxmath}

\usepackage[T1]{fontenc}

\DeclareRobustCommand{\VAN}[3]{#2}
\let\VANthebibliography\thebibliography
\def\thebibliography{\DeclareRobustCommand{\VAN}[3]{##3}\VANthebibliography}
\DeclareRobustCommand{\DE}[3]{#2}
\let\DEthebibliography\thebibliography
\def\thebibliography{\DeclareRobustCommand{\DE}[3]{##3}\DEthebibliography}

\usepackage{graphicx}	
\usepackage{amsmath}	
\usepackage{pdflscape}
\usepackage{rotating}

\PassOptionsToPackage{hyphens}{url}
\usepackage[all]{hypcap}
\usepackage{color}
\definecolor{Brown}{rgb}{0.647,0.165,0.165}
\definecolor{NavyBlue}{rgb}{0.0,0,0.5}
\definecolor{Burgundy}{rgb}{0.5,0.0,0.125}
\definecolor{lime}{rgb}{0.651,0.808,0.224}
\hypersetup{
    breaklinks=true,
    colorlinks=true,
    citecolor={Burgundy},
    linkcolor={NavyBlue},
    urlcolor={Brown}
}
\usepackage{multirow}
\def\apj{Astrophys.\ J.}

\def\mnras{Mon.\ Not.\ R.\ Astron.\ Soc.}
\def\aap{Astron.\ Astrophys.}
\def\apjl{Astrophys.\ J.\ Lett.}

\def\physrep{Phys.\ Rep.}
\def\pre{Phys.\ Rev.\ E}
\def\prl{Phys.\ Rev.\ Lett.}

\def\apjs{ApJS}

\def\pasa{PASA}

\def\nat{Nature}

\def\araa{Ann.\ Rev.\ Astron.\ Astrophys.}

\newcommand{\Mach}{\mathcal{M}}      
 
\newcommand{\dd}{\mathrm{d}}        

\newcommand{\cm}{{\rm cm}}    
\newcommand{\km}{{\rm km}}    
\newcommand{\m}{{\rm m}}      
\newcommand{\pc}{{\rm pc}}     
\newcommand{\kpc}{{\rm kpc}}  

\newcommand{\s}{{\rm s}}      
\newcommand{\Myr}{{\rm Myr}} 

\newcommand{\muG}{\mu{\rm G}} 

\newcommand{\K}{{\rm K}}      
\newcommand{\R}{{\rm R}}      
\newcommand{\mK}{{\rm mK}}      

\newcommand{\rad}{{\rm rad}}

\renewcommand{\ne}{n_{\rm e}}

\newcommand{\RM}{\text{RM}}
\newcommand{\RMe}{\RM_{\rm err}}

\newcommand{\DM}{\text{DM}}
\newcommand{\DMe}{\DM_{\rm err}}
\newcommand{\Lpul}{{L_{\rm pul}}}

\newcommand{\Sol}{\rm Sol}
\newcommand{\Comp}{\rm Comp}
\newcommand{\TIGRESS}{\rm TIGRESS}
\newcommand{\Ekin}{E_{\rm kin}}
\newcommand{\Ekine}{E_{\rm kin, err}}
\newcommand{\Emag}{E_{\rm mag}}
\newcommand{\Emage}{E_{\rm mag, err}}

\newcommand{\dv}{{\delta {v}}}
\newcommand{\dve}{{\dv_{\rm err}}}
\newcommand{\dvt}{{\dv_{\rm turb}}}

\newcommand{\dvth}{{\dv_{\rm ther}}}
\newcommand{\dvtt}{{\dv^{2}_{\rm turb}}}
\newcommand{\dvte}{{\dvt_{\rm, err}}}
\newcommand{\dvHI}{{\delta {v}}_{\rm HI}}
\newcommand{\dvHa}{{\delta {v}}_{{\rm H}\alpha}}
\newcommand{\dvtHI}{{\delta {v}}_{\rm turb, HI}}

\newcommand{\dvtHa}{{\delta {v}}_{{\rm turb, H}\alpha}}
\newcommand{\dvtionHa}{{\delta {v}}_{{\rm turb, ion(H}\alpha)}}
\newcommand{\Blos}{{B_{\rm los}}}
\newcommand{\Bpos}{{B_{\rm pos}}}

\newcommand{\Bloss}{{B^{2}_{\rm los}}}
\newcommand{\Blose}{{\Blos_{\rm, err}}}
\newcommand{\ZHI}{{\rm Zeeman, HI}}
\newcommand{\ZMol}{{\rm Zeeman, Mol}}
\newcommand{\nsamp}{{n_{\rm samp}}}
\newcommand{\GL}{{\rm Gal.\,Lat.}}
\newcommand{\pbeta}{\beta_{\rm plasma}}
\newcommand\Eq[1]{Eq.\,\ref{#1}}
\newcommand\Fig[1]{Fig.~\ref{#1}}
\newcommand\Sec[1]{Sec.~\ref{#1}}
\newcommand\Tab[1]{Table~\ref{#1}}
\newcommand\App[1]{Appendix~\ref{#1}}
\newcommand\rev[1]{{#1}}

\graphicspath{{./}{figs/}}

\title[Magnetic fields in the multiphase ISM]{Magnetic fields in the multiphase interstellar medium of the Milky Way: turbulent kinetic and magnetic energy density relation}

\author[Seta and McClure-Griffiths]{
Amit Seta\thanks{E-mail:  \href{mailto:amit.seta@anu.edu.au}{amit.seta@anu.edu.au}}
and 
N.\ M.\ McClure-Griffiths
\\
Research School of Astronomy and Astrophysics,  Australian National University, Canberra, ACT 2611, Australia\\
}

\date{Accepted XXX. Received YYY; in original form ZZZ}

\pubyear{2025}

\begin{document}
\label{firstpage}
\pagerange{\pageref{firstpage}--\pageref{lastpage}}
\maketitle

\begin{abstract}
Magnetic fields are an important component of the interstellar medium (ISM) of galaxies. The thermal gas in the ISM has a multiphase structure, broadly divided into ionised, atomic, and molecular phases. The connection between the multiphase ISM gas and magnetic field is not known and this makes it difficult to account for their impact on star formation and galaxy evolution. Usually, in star formation studies, a relationship between the gas density, $n$ and magnetic field strength, $B$, is assumed to study magnetic fields' impact. However, this requires the knowledge of the geometry of star-forming regions and ambient magnetic field orientation. Here, we use the Zeeman magnetic field measurements from the literature for the atomic and molecular ISM and supplement the magnetic field estimates in the ionised ISM using pulsar observations to find a relation between the turbulent kinetic, $E_{\rm kin}$, and magnetic, $E_{\rm mag}$, energy densities. Across all three phases and over a large range of densities ($10^{-3}\,{\rm cm}^{-3} \lesssim n  \lesssim 10^{7}\,{\rm cm}^{-3}$), we find $\Emag \propto \Ekin$. Furthermore, we use phase-wise probability density functions of density, magnetic fields, and turbulent velocities to show that the magnetic field fluctuations are controlled by both density and turbulent velocity fluctuations. This work demonstrates that a combination of both the density and turbulent velocity determines magnetic fields in the ISM.
\end{abstract}




\section{Introduction} \label{sec:int}
The interstellar medium (ISM) of star-forming galaxies is a dynamic medium between stars consisting of thermal gas, dust, radiation, turbulence, magnetic fields, and cosmic rays. Magnetic fields are an energetically important component of the ISM as they play a crucial role in star formation \citep{MestelS1956, KrumholzF2019, PattleEA2023}, propagation of cosmic rays \citep{Cesarsky1980, Zweibel2017, RuszkowskiP2023}, determining properties of galactic outflows \citep{VeilleuxEA2005, VoortEA2021, ThompsonH2024}, and might also play a significant role in the evolution of galaxies \citep{NaabO2017, Martin-AlvarezEA2020, PakmorEA2024}. Despite their significance, the ISM's thermal gas-magnetic field connection is poorly understood. This makes it difficult to study the role of magnetic fields in star formation and galaxy evolution -- two outstanding problems in modern astrophysics.

The matter and energy in galaxies cycles between the stars and the ISM. Stars are continuously formed from the cold, dense, molecular ISM, where the self-gravity of gas leads to a collapse. Then the stars through stellar outflows and, primarily for massive stars, supernova explosions, return energy and material back into the ISM generating turbulence and the hot, ionised ISM gas. The hot, ionised gas cools \citep{SutherlandD1993} to form the warm, atomic ISM. The atoms combine to form molecules creating the molecular ISM, which again could potentially form stars. Thus, the gas in galaxies cycles through these multiple phases and in a dynamic equilibrium state, it gives rise to a multiphase structure \citep{FieldEA1969, McKeeO1977, Cox2005}. More generally, the ISM gas can be in various forms: molecular (cold), atomic (cold and warm), and ionised (warm and hot) phases \citep[also see][]{DickeyL1990, WolfireEA1995, WolfireEA2003, KalberlaK2009, Ferriere2020, SaintongeC2022, McClure-GriffithsEA2023}. The properties of several ISM components, most notably, turbulence, magnetic fields, and cosmic rays vary with the ISM phase. In this work, we primarily study the multiphase gas-magnetic field connection.

Between the ISM phases (ionised, atomic, and molecular), the gas number density, $n$, varies over a huge range, from $10^{-3}\,\cm^{-3}$ in the ionised ISM to up to $10^{7}\,\cm^{-3}$ in the molecular ISM \citep[see Table 1 in][]{Ferriere2020}. In collapsing molecular clouds, magnetic fields amplify due to compression. A relationship between $n$ and magnetic field strength, $B$, is usually assumed in such a scenario to determine the dynamical importance of magnetic fields on the formation of stars. For example, in the case of isotropic spherical collapse, $B \propto n^{2/3}$ \citep{Mestel1966} and for collapse parallel to the magnetic field, $B \propto n^{1/2}$ \citep{Mouschovias1976a, Mouschovias1976b}. Generally, depending on the geometry of the collapsing cloud and the orientation of the magnetic field relative to the collapsing direction, the $B$ -- $n$ relation could vary \citep[see Fig. 1 in][]{TritsisEA2015}. Using Zeeman measurements of magnetic field strengths in the atomic (HI) and molecular (OH and CN) ISM, \citet{CrutcherEA2010} showed that $B$ -- $n$ relation is a broken power-law function with $B \propto n^{0}$ below the critical density $n_{\rm crit} \approx 300\,\cm^{-3}$ and $B \propto n^{2/3}$ above that. \citet{TritsisEA2015} reanalysed the same data and, for the molecular ISM, concluded $B \propto n^{1/2}$ instead of $B \propto n^{2/3}$. Including magnetic fields in the ionised phase of HII regions with these Zeeman measurements, \citet{Harvey-SmithEA2011} found $B \propto n^{0.11}$ below $n_{\rm crit} \approx 480\,\cm^{-3}$ and $B \propto n^{0.66}$ for densities above it. \rev{Using far-infrared and HI data, \citet{KalberlaH2023} found $B \propto n^{0.58}$ for $0.1\,\cm^{-3} \lesssim n \lesssim 100\,\cm^{-3}$.} More recently, \citet{WhitworthEA2024}, combined these Zeeman measurements with magnetic fields from dust polarisation observations \citep{PattleEA2023} and concluded $B \propto n^{(0.27 \pm 0.017)}$ for $n \le n_{\rm crit} \approx 924^{+154}_{-144}\,\cm^{-3}$ and $B \propto n^{(0.54 \pm 0.18)}$ for $n > n_{\rm crit}$. In this paper, besides the atomic and molecular ISM, we also explore magnetic fields in the ionised ISM using pulsar observations.

A variety of magnetohydrodynamic turbulence simulations show a huge scatter around and differences with the expected $B$ -- $n$ relations (e.g.,~see Fig.~5 in \citealt{SetaF2022}, Fig.~1 in \citealt{PonnadaEA2022}, Fig.~7 in \citealt{HuEA2023}, Fig.~7 in \citealt{KonstantinouEA2024}, Fig.~5 -- 6 in \citealt{HeR2024}, and Fig.~6 -- 14 in \citealt{WhitworthEA2024}) demonstrating that, in general, the magnetic field does not necessarily depend only on the gas density. The prevailing ISM turbulence, especially in the ionised and atomic ISM, also further complicates the applicability of the $B$ -- $n$ relations (which are largely based on systematic compression or collapse scenarios). Turbulence in the ISM is driven on a range of scales by different processes, from stellar outflows at sub-$\pc$ scales to supernova explosions at $\approx 100\,\pc$ \citep{MacLowK2004, ElmegreenS2004, ScaloE2004} to gravitational instabilities at even larger $\kpc$ scales \citep{KrumholzB2016, KrumholzEA2018}. The properties of turbulence also vary with the ISM phase: turbulence is supersonic in the cold (atomic and molecular) ISM and transonic/subsonic in the warm (ionised and atomic) and hot (ionised) ISM \citep{Larson1981, HeilesT2003II, GaenslarEA2011, MurrayEA2015, NguyenEA2019, MarchalM2021, FederrathEA2021}. Both the density and turbulent velocity in the ISM phases could be together characterised using the turbulent kinetic energy, $\Ekin = 0.5\, n\,\dvtt$, where $\dvt$ is the turbulent velocity. Using $\Ekin$, in this work, we aim to determine a relation that captures the multiphase gas-magnetic field connection better than the existing $B$ -- $n$ relations and also factors in the turbulence scenario.

In addition to introducing density and velocity fluctuations, ISM turbulence also affects magnetic field properties. Magnetohydrodynamic turbulence simulations, which start with initially weak magnetic fields, show magnetic field amplification and, once the magnetic field achieves a statistically steady state, the amplified magnetic energy density, $\Emag = B^{2} / 8 \pi$, is proportional to $\Ekin$ (this need not necessarily be the case for simulations that start with dynamically strong fields). This is due to a dynamo process, which is the conversion of turbulent kinetic energy into magnetic energy \citep{Kazantsev1968, ZeldovichEA1984,  RuzmaikinEA1988, KulsrudA1992, BeckEA1996, BrandenburgS2005, Rincon2019, ShukurovS2021, Schekochihin2022}. The dynamo amplifies weak seed magnetic fields in protogalaxies and the early Universe \citep[$\sim 10^{-10}\,\muG$; see][]{KulsrudEA1997b, Subramanian2016, SetaF2020} to roughly equipartition field strengths (i.e., $\Emag \approx \Ekin,\,\sim1$ -- $10\,\muG$) and maintains it against decay and dissipation. The equipartition field strengths are also observed in present-day \citep{Beck2016} and relatively young  \citep{BernetEA2008, MaoEA2017, SetaEA2021, ShahS2021, MahonyEA2022} galaxies. This process has been extensively studied using a variety of numerical simulations \citep{MeneguzziEA1981, SchekochihinEA2004, HaugenEA2004, FederrathEA2011, SchoberEA2012c, GentEA2013, FederrathEA2014, Federrath2016, RiederT2017a, SetaEA2020, SetaF2021dyn, AchikanathChirakkaraEA2021, SetaF2022, GentEA2023, BrandenburgEA2023, SurS2024, Korpi-LaggEA2024}. For all of these simulations, in the statistically steady state, $\Emag \propto \Ekin$. Although the level of energy equipartition ($\Emag/\Ekin)$ may depend on the ISM phase \citep[$\Emag/\Ekin$ is roughly an order of magnitude higher in the warm phase in comparision to the cold phase, see][for further details]{SetaF2022}. In this paper, using multiple Milky Way observations, we show that $\Emag \propto \Ekin$ holds across several ISM phases over a large range of densities and is a physically more concrete alternative to the popular $B$ -- $n$ relations.

The paper is organised as follows. In \Sec{sec:met}, we present the observational datasets and methods used to derive $\Emag$ and $\Ekin$. The $\Emag$ -- $\Ekin$ relation  is discussed in \Sec{sec:res}. In \Sec{sec:dis}, we further discuss the assumptions in and impact of the derived $\Emag$ -- $\Ekin$ relation. Finally, we summarise and conclude our work in \Sec{sec:con}.

\section{Data and Method} \label{sec:met}
In this paper, we seek to compare data that probe the magnetic field energy density, $\Emag$, and the kinetic energy density, $\Ekin$, across molecular, atomic, and ionised gas phases.  To do this, we compile observations of the Zeeman effect in atomic and molecular gas, which give localised measurements of the line-of-sight magnetic field strength and turbulent velocity widths, along with estimates of the gas density.  We augment these with estimates of the mean line-of-sight gas density and magnetic field strength from pulsar observations, combined with estimates of the turbulent velocity from both ionised and atomic gas tracers.  Inherent in our analysis is the assumption that when considering population-based statistics, the line-of-sight mean properties for ionised gas can be compared with localised measurements, like those from Zeeman (further discussed in \Sec{sec:pul}).  In this section, we describe the data sources and our method, in \Sec{sec:res}, we present our results, and in \Sec{sec:dis}, we justify the validity of the assumptions that are inherent in comparing these different types of data.

\subsection{Zeeman data for the atomic and molecular ISM} \label{sec:zdata}
The Zeeman data for both the atomic (HI) and molecular (OH and CN) ISM is taken from Table~1 in \citet{CrutcherEA2010}. This includes the line-of-sight magnetic fields, $\Blos$, the estimated uncertainty in $\Blos$, $\Blose$, and the gas density, $n$, for $137$ sources. We divide the entire data set into two subsets: one which probes the atomic ISM (referred to as the $\ZHI$ dataset) and the other which probes the molecular ISM (referred to as the $\ZMol$ dataset). 

The velocity widths, $\dv$, are also obtained differently for $\ZHI$ and $\ZMol$ datasets. For the atomic ISM, the Stokes I spectrum (total intensity vs. velocity) of the HI line is fitted with Gaussian distributions \citep{HeilesT2003I} to determine the magnetic field strength \citep[e.g.~see Sec. 2.8 in][]{HeilesT2004III}, which gives the velocity line widths, $\dv$, and associated uncertainties, $\dve$, from the fitting. The $\dv$ and $\dve$ corresponding to the sources in the $\ZHI$ dataset are taken from Table 1 in \citet{HeilesT2004III}. For the molecular ISM, the numerical derivative of the Stokes I spectrum is fitted to the Stokes V (circular polarisation) spectrum to obtain the magnetic field strength \citep[e.g.~see Eq.~3 in][]{CrutcherK2019} and $\dv$s are known. For the $\ZMol$ dataset, the corresponding $\dv$s are taken from the literature \citep[][and references there in]{Crutcher1999, TrolandC2008, FalgaroneEA2008} and, for each source, the $\dve$ is taken to be $5\%$ of their corresponding line-width, $\dv$ (Crutcher, Richard M., private communication).

After compiling the data ($n, \Blos, \Blose, \dv, \dve$) for all of the $137$ sources in Table 1 of \citet{CrutcherEA2010}, we discard sources with $\Blos=0$ (one component of the source 3C 348 in the $\ZHI$ dataset) and $\dve=0$ (one component of two sources, 3C 225b and 3C 237, in the $\ZHI$ dataset). This leaves us with a total of $134$ sources in our sample, $66$ in the $\ZHI$ dataset and $68$ in the $\ZMol$ dataset.

\subsection{Pulsar data for the ionised ISM} \label{sec:pdata}

We use pulsar observations to probe the gas (thermal electron) number density and magnetic fields in the ionised ISM. Due to the thermal electron density, $\ne$, along the path between the pulsar and us, the pulsed signals are delayed to lower frequencies and the observed dispersion in the pulse is quantified in terms of the dispersion measure, $\DM$, given by
\begin{align} \label{eq:dm}
\frac{\DM}{\pc\,\cm^{-3}} = \int_{\Lpul/\pc} \frac{\ne}{\cm^{-3}} \dd \left(\frac{l}{\pc}\right),
\end{align}
where $\Lpul$ is the distance from the pulsar to us. Also, the linearly polarised emission from the pulsar undergoes Faraday rotation due to $\ne$ and line-of-sight magnetic fields, $\Blos$, along the path, which is the rotation of the polarisation plane as a function of the observing wavelength, $\lambda$, as $\delta \psi = \RM\,\lambda^{2}$, where $\delta \psi$ is the change in polarisation angle between the source and observer and $\RM$ is the rotation measure, given by
\begin{align}\label{eq:rm}
\frac{\RM}{\rad\,\m^{-2}} = 0.812\,\int_{\Lpul/\pc} \frac{\ne}{\cm^{-3}} \frac{\Blos}{\muG} \dd \left(\frac{l}{\pc}\right).
\end{align}

Knowing the distance to the pulsar, $\Lpul$, the average thermal electron density, $\langle \ne \rangle$, can be estimated from the observed $\DM$ as
\begin{align} \label{eq:meanne}
\frac{\langle \ne \rangle}{\cm^{-3}} = \frac{\DM}{\pc\,\cm^{-3}}\left(\frac{\Lpul}{\pc}\right)^{-1}.
\end{align}
Furthermore, from $\RM$ (\Eq{eq:rm}) and $\DM$ (\Eq{eq:dm}), the average $\Blos$ can be obtained as 
\begin{align} \label{eq:avgbpar}
\frac{\langle \Blos \rangle}{\muG} = 1.232~\frac{\RM}{\rad\,\m^{-2}} \left(\frac{\DM}{\pc\,\cm^{-3}}\right)^{-1}.
\end{align}
Both the derived quantities (\Eq{eq:meanne} and \Eq{eq:avgbpar}) from $\RM$ and $\DM$ observations have been extensively used to study the properties of the ionised ISM and Galactic magnetic fields \citep{Smith1968, HewishEA1968, Manchester1972, Manchester1974, LyneS1989, RandK1989, HanQ1994, HanEA1999, IndraniD1999, MitraEA2003, BerkhuijsenEA2006, HanEA2004, HanEA2006, BerkhuijsenF2008, GaenslerEA2008, HanEA2018, YaoEA2017, SobeyEA2019, LeeEA2024}. Note that using \Eq{eq:avgbpar} to estimate the average line-of-sight magnetic field assumes that the thermal electron density and magnetic fields are uncorrelated \citep{BecKEA2003} and this assumption holds for larger $\approx \kpc$ distances to pulsars \citep{SetaF2021pul}.

We use $\RM$ and $\DM$ observations from the ATNF pulsar catalogue \citep[][\href{https://www.atnf.csiro.au/research/pulsar/psrcat}{https://www.atnf.csiro.au/research/pulsar/psrcat}, version 2.4.0]{ManchesterEA2005} for pulsars with only independently determined distances (\texttt{Dist$_{\texttt A}$} in the catalogue, referred to as $\Lpul$, $271$ pulsars in total). From this dataset, we remove pulsars with $\RM=0$, the error in $\RM$, $\RMe=0$, $\DM=0$, and the error in $\DM$, $\DMe=0$. This leaves us with $237$ pulsars. Additionally, we choose pulsars with distances, $\Lpul \le 10\,\kpc$, which further reduces the dataset to $212$ pulsars. From this dataset, we compute $\langle \ne \rangle$ and $\langle \Blos \rangle$ using \Eq{eq:meanne} and \Eq{eq:avgbpar}, respectively. For the ionised ISM, as probed by the pulsar sample, we assume the gas with $n \approx \langle \ne \rangle$ hosts $\Blos \approx \langle \Blos \rangle$ (this assumption is thoroughly discussed in \Sec{sec:pul}).

\subsection{Velocity widths for the ionised ISM, $\dvHa$ and $\dvHI$} \label{sec:vdata}
The pulsar DM and RM measurements do not have any inherent mechanism for estimating the velocity width.  To compute velocity widths for the ionised ISM probed by the pulsars, we use spectral lines tracing both the ionised (H$\alpha$) and atomic (HI) hydrogen along pulsar sight lines. The H$\alpha$ spectra for pulsar locations are sourced from the Wisconsin H$\alpha$ Mapper Sky Survey \citep[WHAM survey,][\href{https://www.astro.wisc.edu/research/research-areas/galactic-astronomy/wham/wham-sky-survey}{https://www.astro.wisc.edu/research/research-areas/galactic-astronomy/wham/wham-sky-survey}]{HaffnerEA2003, HaffnerEA2010} and has an angular and spectral resolution of $1\,\deg$ and $12\,\km\,\s^{-1}$, respectively. The HI emission spectra (data sourced from \href{https://www.astro.uni-bonn.de/hisurvey/AllSky_gauss}{https://www.astro.uni-bonn.de/hisurvey/AllSky\_gauss}), depending on the location of the pulsar, are either taken from the Effelsberg - Bonn HI Survey \citep[EBHIS,][]{WinkelEA2016} or Parkes Galactic All-Sky Survey \citep[GASS,][]{McClure-GriffithsEA2009, KalberlaEA2010}. The EBHIS dataset has an angular and spectrum resolution of $10.8$ arcmin and $1.29\,\km\,\s^{-1}$, respectively, and the GASS dataset has those as $16.2$ arcmin and $0.82\,\km\,\s^{-1}$. The HI spectra have a significantly better resolution, especially spectral resolution, than the H$\alpha$ dataset.

From the H$\alpha$ and HI data along pulsar sightlines for the Milky Way (local standard of rest velocities within $\pm\,50\,\km\,\s^{-1}$, see \Sec{sec:miss} for a discussion on this choice), we estimate the velocity widths, $\dv$, by computing the square root of the second moment of the spectral line, $m_{2}$, as
\begin{align} \label{eq:m2}
\dv = m_{2}^{1/2} = \left(\frac{\sum_{i=1}^{n_{\rm chan}} {\rm amp}_{i} (v_{i} - m_{1})^{2}}{\sum_{i=1}^{n_{\rm chan}}{\rm amp}_{i}}\right)^{1/2},
\end{align}
where $n_{\rm chan}$ is the number of velocity channels, $v_{i}$ is the velocity in each channel, ${\rm amp}_{i}$ is the amplitude of the line in each channel (in Rayleighs, $\R$, for the H$\alpha$ spectra and in Kelvin, $\K$, for the HI spectra), and $m_{1} = {\sum_{i=1}^{n_{\rm chan}} {\rm amp}_{i} v_{i}} / {\sum_{i=1}^{n_{\rm chan}}{\rm amp}_{i}}$ is the first moment. For this computation, the spectra are thresholded at three times the noise level to avoid contributions from spurious peaks (especially at higher velocities), which is a location-dependent quantity in the WHAM survey for the H$\alpha$ spectra \citep[average level of $0.15\,\R$, see][]{HaffnerEA2003} and $\approx 100\,\mK$ for the HI spectra \citep[see Table 1 in][]{HI4PI2016}. \rev{Representative H$\alpha$ and HI spectra, along with the computed $m_{1}$ and $m_{2}$, are shown in \Fig{fig:spectra} and further discussed in \App{sec:spectra}.} For both the H$\alpha$ and HI datasets, the error in $\dv$, $\dve$, is estimated via a Monte Carlo-type method. Random velocities are drawn from each channel (keeping channel width the same) for random amplitudes taken from the range (measurement $\pm$ three times the observed noise level) 5000 times to compute a distribution of $\dv$ for each pulsar sightline using \Eq{eq:m2}. The difference between $84^{\rm th}$ and $16^{\rm th}$ percentile of this distribution is taken as $\dve$.

\subsection{Construction of $\dvt$ from $\dv$} \label{sec:dvt}
Once the velocity widths for the pulsars (both $\dvHa$ and $\dvHI$), $\ZHI$, and $\ZMol$ datasets are known, we need to estimate the turbulent velocity, $\dvt$, by removing the contribution from the thermal speed, $\dvth$. Knowing the Mach number of the turbulence ($\Mach = \dvt/\dvth$) and $\dv = (\dv^{2}_{\rm ther} + \dv^{2}_{\rm turb})^{1/2}$, we get 
\begin{align} \label{eq:dvt}
\dvt = \dv \left(\frac{1}{\Mach^{2}} + 1\right)^{-1/2}.
\end{align}

Based on the literature, we assume that the Mach number in the ionised ISM, $\Mach_{\rm ion} \approx 1$, for both the H$\alpha$ \citep{HillEA2008, GaenslarEA2011} and HI \citep[][\rev{also, see a discussion in \Sec{sec:miss} about Mach numbers in the atomic ISM from HI spectra}]{HeilesT2003II,MarchalM2021} datasets. In the atomic ISM, $\Mach_{\rm atm} \approx 1$, for the $\ZHI$ dataset \citep{HeilesT2003II,MarchalM2021} and in the molecular ISM, $\Mach_{\rm mol} \approx 5$ for the $\ZMol$ dataset \citep{SchneiderEA2013}. Assuming these Mach numbers and from the $\dv$, $\dvt$ are computed for all three (pulsars, $\ZHI$, and $\ZMol$) datasets using \Eq{eq:dvt}. The error in $\dv$ is propagated to compute the error in $\dvt$, $\dvte$. 

Once $\dvt$ is known, $\Ekin = 0.5\,n\,\dvtt$ and $\Emag =  \Bloss / 8 \pi$ is computed for all three datasets and the uncertainties in $\dvt$ and $\Blos$ are propagated to estimate uncertainties, $\Ekine$ and $\Emage$ ($n$ has a significantly low error compared to $\dvt$, also for the pulsar samples because of the comparatively lower relative level of uncertainties in $\DM$ compared to both the H$\alpha$ and HI spectra, \rev{see \Sec{sec:mres} for further discussion}). \rev{From now on in the paper, $\Emag$ always refers to magnetic energy only along the line-of-sight and thus is the lower limit of the total magnetic energy.}

\section{Results} \label{sec:res}
In this section, we use the entire data, consisting of three datasets: pulsars probing the magnetic fields in the ionised ISM (ion), $\ZHI$ probing the magnetic fields in the atomic ISM (atm), and $\ZMol$ probing the magnetic fields in the molecular ISM (mol) to show $\Emag \propto \Ekin$ holds across all three phases (and equivalently over a large range of gas densities). We also discuss that the magnetic fields depend on both the gas density and turbulent velocity, not just the gas density as usually proposed by the $B$ -- $n$ relations.

\subsection{$\Ekin \propto \Emag$ instead of $B \propto n^{\rm factor}$} \label{sec:mres}

\begin{figure}
\includegraphics[width=\columnwidth]{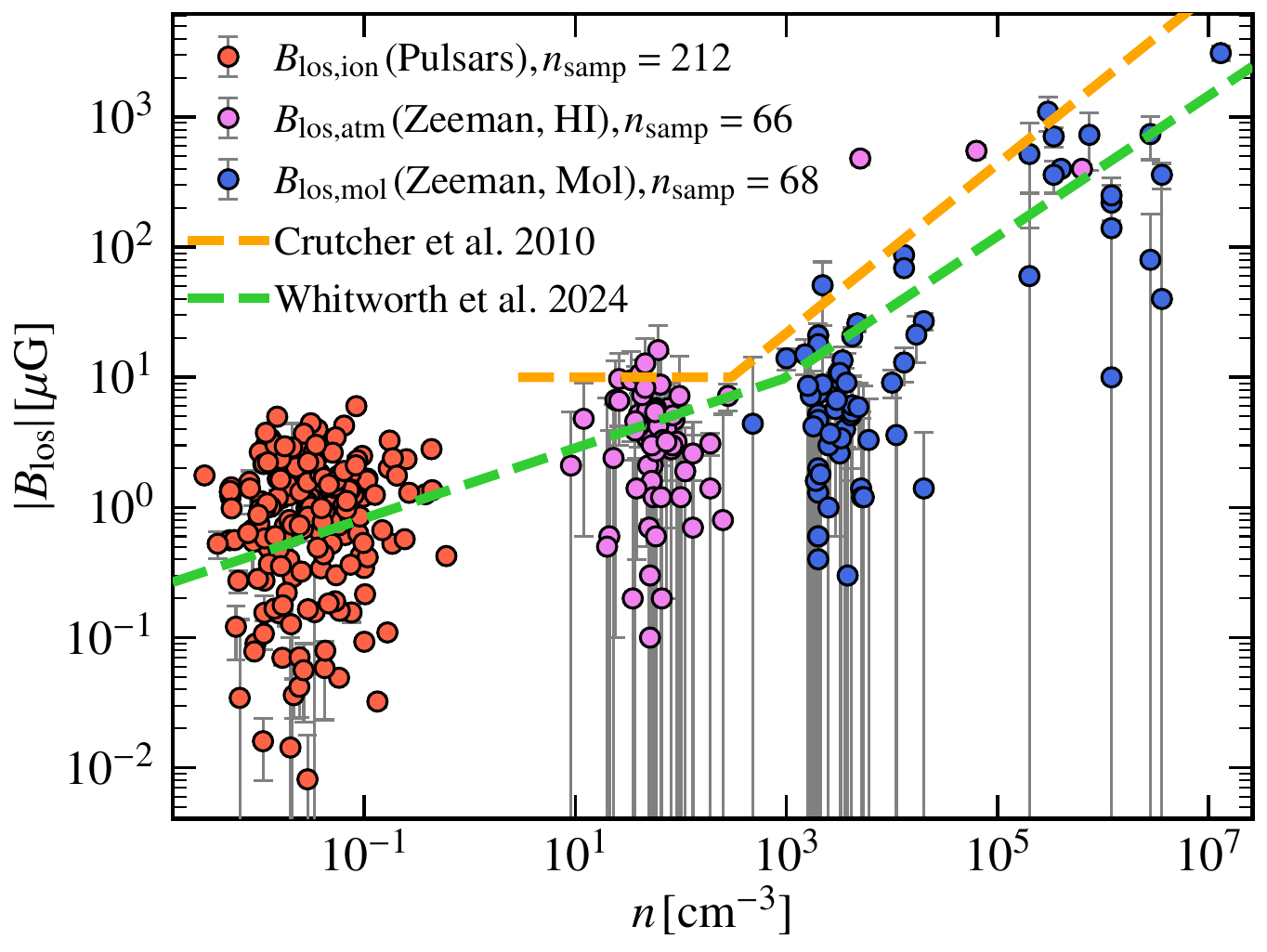}
\caption{
The relationship between density, $n\,[\cm^{-3}]$, and line-of-sight magnetic fields, $\Blos\,[\muG]$, for Zeeman measurements probing atomic ($\ZHI$) and molecular ($\ZMol$) ISM and for the ionised ISM added using pulsars in this work. $\nsamp$ represents the number of data points in each sample. The dashed orange and green lines show broken power-law relations taken from the literature. The \citealt{CrutcherEA2010} line shows $10\,\muG$ for $n \lesssim n_{\rm crit} \approx 300\,\cm^{-3}$ and $10\,\muG (n/n_{\rm crit})^{2/3}$ for $n > n_{\rm crit}$ and the \citealt{WhitworthEA2024} line shows $9.685\,\muG\,(n/n_{\rm crit})^{0.27}$ for $n \lesssim n_{\rm crit} \approx 924\,\cm^{-3}$ and $9.685\,\muG (n/n_{\rm crit})^{0.54}$ for $n > n_{\rm crit}$. The \citealt{CrutcherEA2010} relation is only shown across $\ZHI$ and $\ZMol$ datasets as it was derived only for those range of densities whereas the \citealt{WhitworthEA2024} relation was derived using a much larger range of densities (without studying the ionised phase though). The \citealt{WhitworthEA2024} relation seems to better capture the data, also for the ionised ISM phase which was not used to derive that relation.
}
\label{fig:blosn}
\end{figure}

First, in \Fig{fig:blosn}, we show the familiar $\Blos$ vs. $n$ plot, especially including the pulsar data probing the ionised ISM in addition to Zeeman measurements. The pulsar data occupies significantly lower values of density ($3 \times 10^{-3}\,\cm^{-3} \lesssim n  \lesssim 6 \times 10^{-1}\,\cm^{-3}$) for the diffuse ionised ISM, which also hosts statistically lower values of magnetic field strengths (see \Fig{fig:pdf}~(b) and \Sec{sec:pdf} for further details). Based on the range of densities used to derive $B$ -- $n$ relations in the literature, we also show the broken power-law trend from \citet{CrutcherEA2010} for the atomic and molecular data and that from \citet{WhitworthEA2024} for the entire range of densities. The relationship in \citet{WhitworthEA2024} also seem to capture the ionised ISM well (note that they did not use the ionised ISM observations to derive the relation, they used a combination of dust polarisation and Zeeman observations).

\begin{figure*}
\centering
\includegraphics[width=13.5cm, height=10cm]{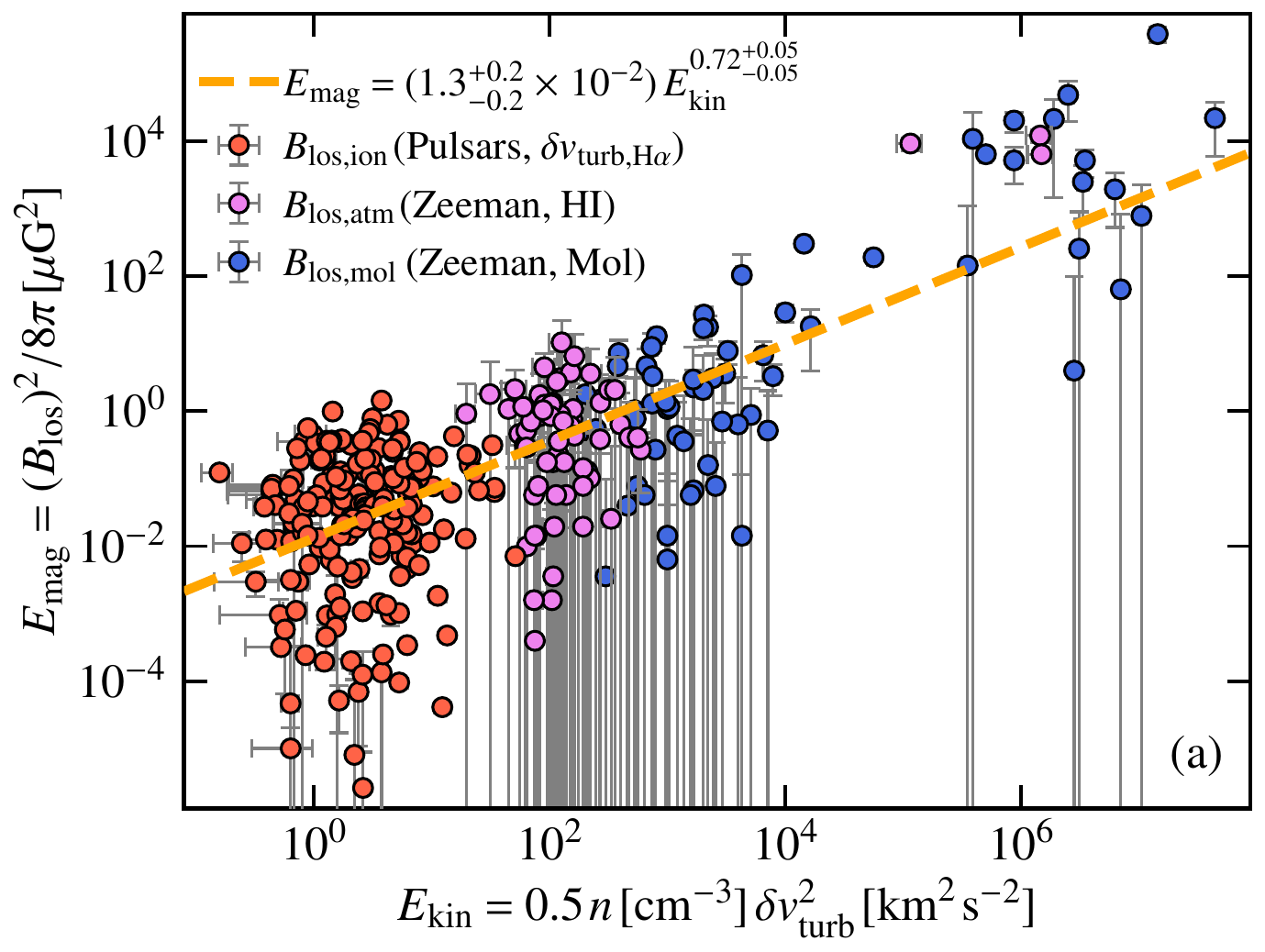} \\ \vspace{0.5cm}
\includegraphics[width=13.5cm, height=10cm]{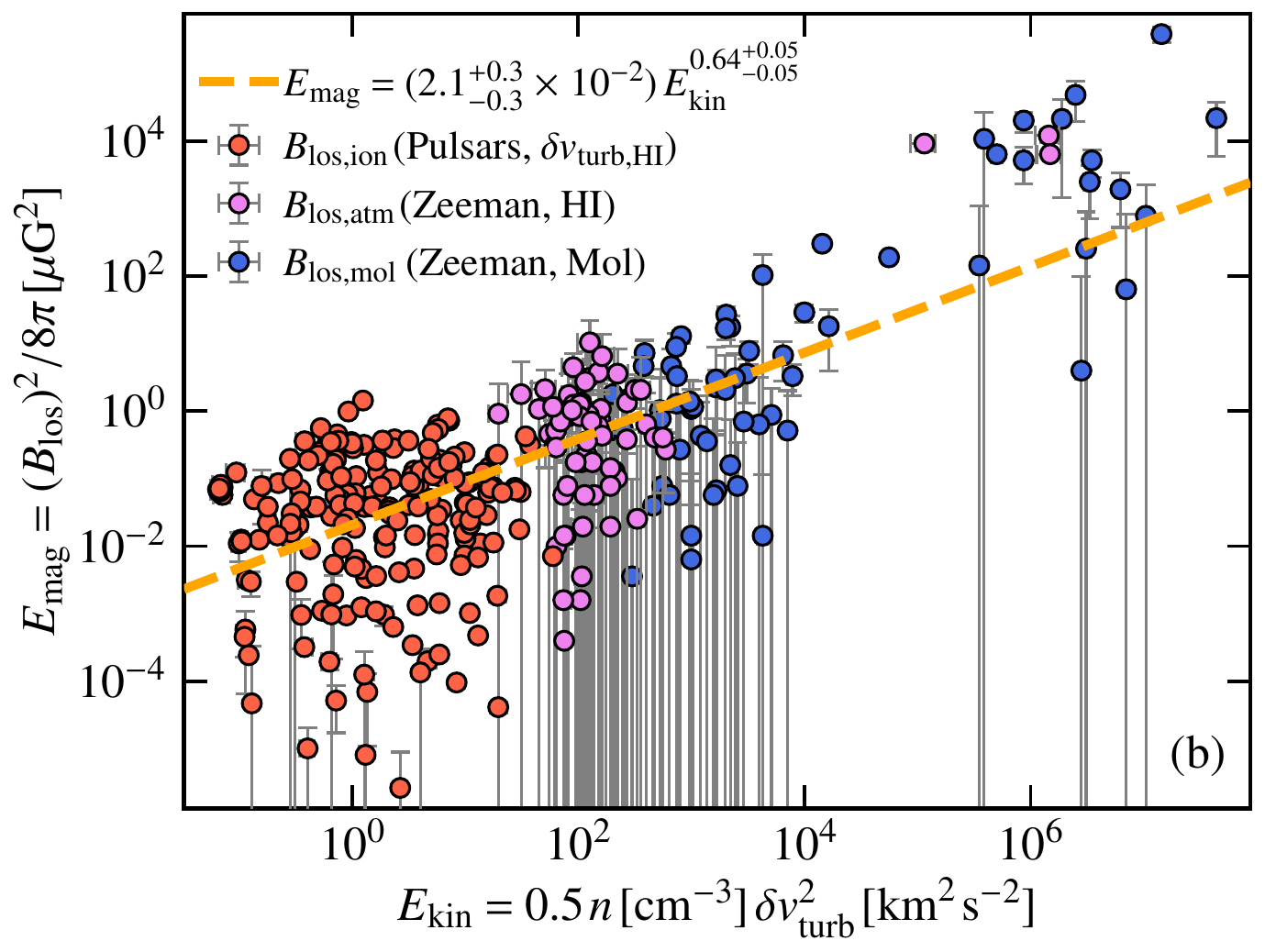}
\caption{The turbulent kinetic, $\Ekin$, and magnetic, $\Emag$, energy densities for the ionised (red, probed using pulsars), atomic (magenta, probed with $\ZHI$), and molecular (blue, probed with $\ZMol$) ISM with a power-law fit (dashed orange line) for ionised gas turbulent velocities, $\dvt$, computed using H$\alpha$ (panel~(a), $\dvtHa$) and HI (panel~(b), $\dvtHI$) spectra. The uncertainties in $\Ekin$ values are higher for the H$\alpha$ case due to a larger velocity channel width of $12\,\km\,\s^{-1}$ for the H$\alpha$ spectra as compared to $\approx 1\,\km\,\s^{-1}$ for the HI spectra. The exponent of the fitted power-law is slightly higher for the H$\alpha$~(a) case in comparision to the HI~(b). But, for both cases, $\Emag \propto \Ekin$, over all three phases and by extension a large range of densities ($10^{-3}\,\cm^{-3} \lesssim n \lesssim 10^{7}\,\cm^{-3}$, see \Fig{fig:blosn}). }
\label{fig:emagekin}
\end{figure*}

Now, in \Fig{fig:emagekin}, we show the relationship between $\Emag$ and $\Ekin$ for the entire dataset, focussing on $\dvt$ for the ionised ISM derived using the H$\alpha$ and HI spectra in \Fig{fig:emagekin}~(a) and \Fig{fig:emagekin}~(b), respectively. For both cases, over the entire range of densities ($10^{-3}\,\cm^{-3} \lesssim n  \lesssim 10^{7}\,\cm^{-3}$), $\Emag \propto \Ekin$. The data is fitted using a Monte Carlo-type method, where 5000 random samples are drawn in the $xy$-range ($\Ekin \pm \Ekine,\,\Emag \pm \Emage$) and fitted with a power-law, $\Emag = \mathcal{C}_{E}\,\Ekin^{\alpha_{E}}$ using $\texttt{LMFIT}$ \citep{NewvilleEA2015}. Then the coefficient, $\mathcal{C}_{E}$, and exponent, $\alpha_{E}$, are estimated from the distributions to be the $50^{\rm th}$ percentiles and the corresponding errors are calculated as the difference between the $84^{\rm th}$ and $16^{\rm th}$ percentiles. For both the H$\alpha$ ($\dvtHa$) and HI ($\dvtHI$) datasets, the coefficients and exponents (reported in the legends of \Fig{fig:emagekin}) are similar, with the coefficient being slightly higher when $\dvtHI$ is used and exponent being slightly higher on using $\dvtHa$. The uncertainties in $\Ekin$ with $\dvtHa$ are higher than that for $\dvtHI$ (due to the smaller velocity channel width of the HI spectra, $\approx 1\,\km\,\s^{-1}$, compared to the H$\alpha$ spectra, $12\,\km\,\s^{-1}$) but the fitted parameters and associated errors are more driven by the significantly higher uncertainties in the $\Emag$ values for both cases. Given the huge spread in $\Emag$, it is difficult to determine the exact fraction of the turbulent kinetic energy being converted to magnetic energy (this fraction might also strongly depend on the ISM phase) but $\Emag \propto \Ekin$ holds true over all three phases (\rev{see \App{sec:Zeeman} for the analysis with just the Zeeman measurements and} \App{sec:emagekindcf} for the analysis which also includes plane-of-sky observations from dust polarisation measurements)\footnote{\rev{The uncertainties in the estimated density can significantly affect the derived  $B$ -- $n$  relations \citep{TritsisEA2015, JiangEA2020}. To assess the impact of density uncertainties on the $\Emag$ -- $\Ekin$ relation, we performed a numerical experiment in which we randomly assigned density uncertainties uniformly within a range of $10\%$ to $300\%$ of the respective density values for each point. We then reconstructed $\Ekine$ and fit the $\Emag$ -- $\Ekin$ relation, incorporating these density uncertainties. The differences in the fitted parameters, compared to those in \Fig{fig:emagekin}, are negligible, as the uncertainties in the magnetic fields and turbulent velocities primarily dominate the overall uncertainties in the fit parameters.}}. This also implies that the magnetic fields in the multiphase ISM are determined by a combination of density and turbulent velocity (via $\Ekin$) and do not only depend on the density as might be inferred from $B$ -- $n$ relations.

\subsection{Probability distribution functions (PDFs) of density, magnetic fields, and turbulent velocities} \label{sec:pdf}

\begin{figure*}
\includegraphics[width=\columnwidth]{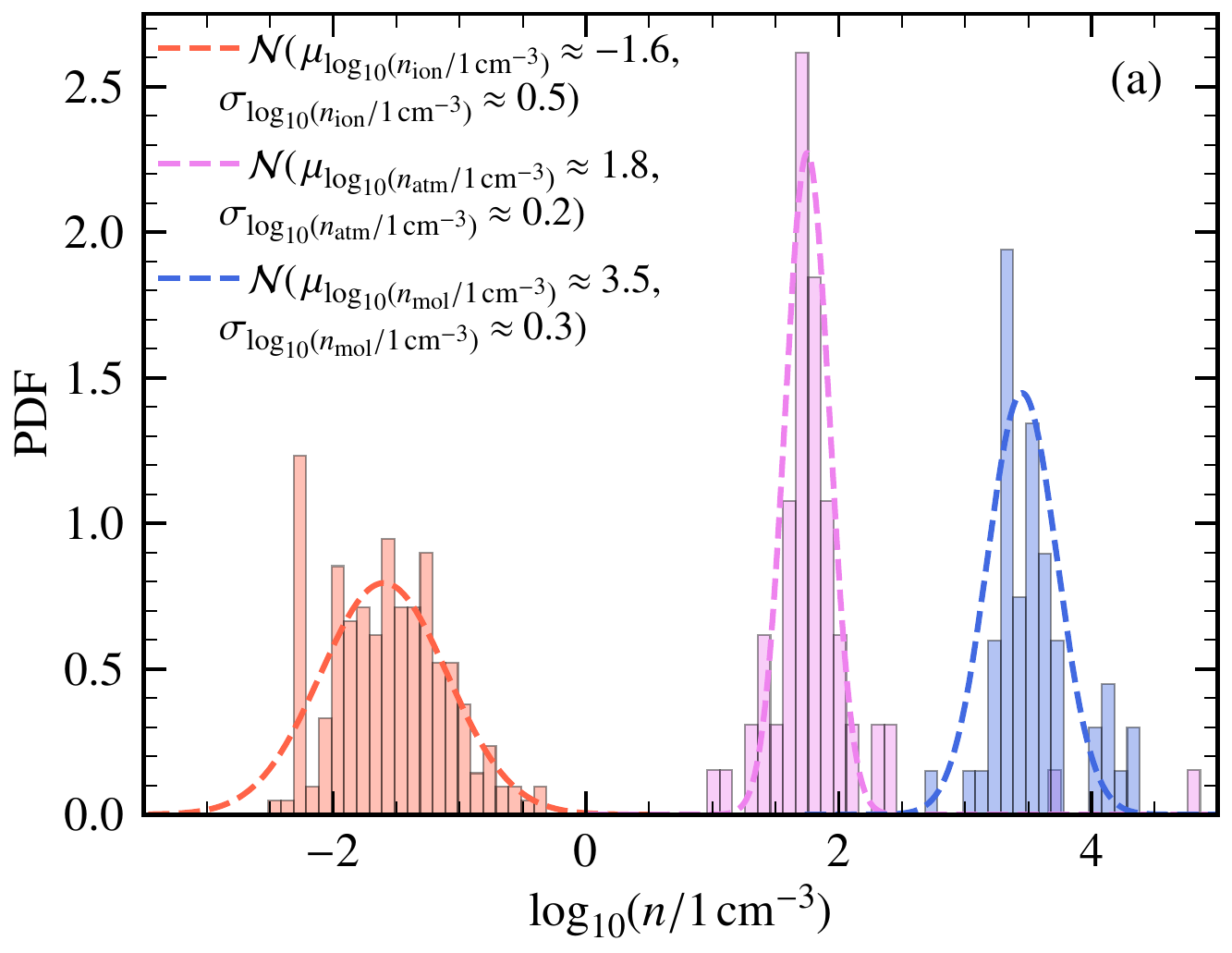}  \hspace{0.5cm}
\includegraphics[width=\columnwidth]{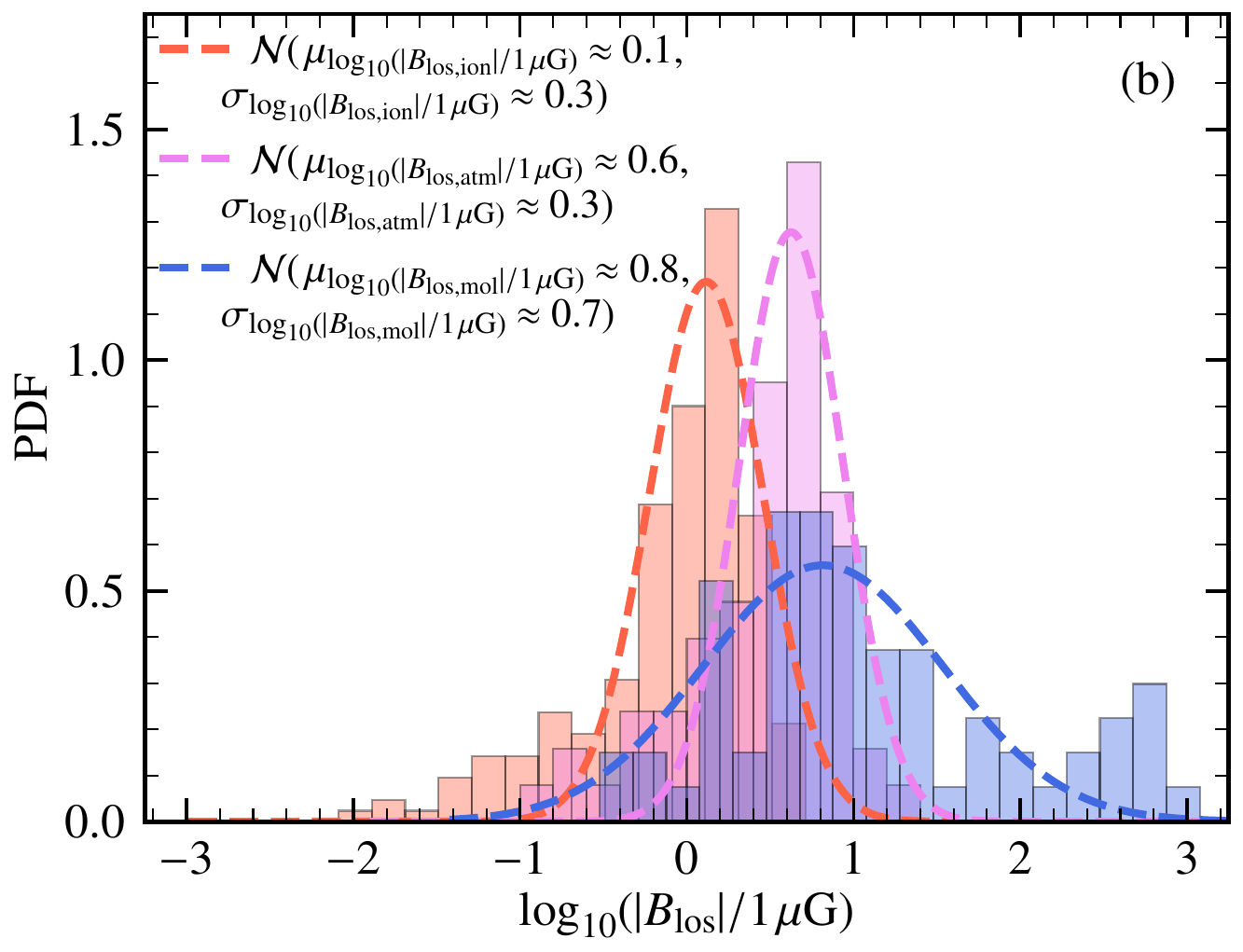} \\
\includegraphics[width=\columnwidth]{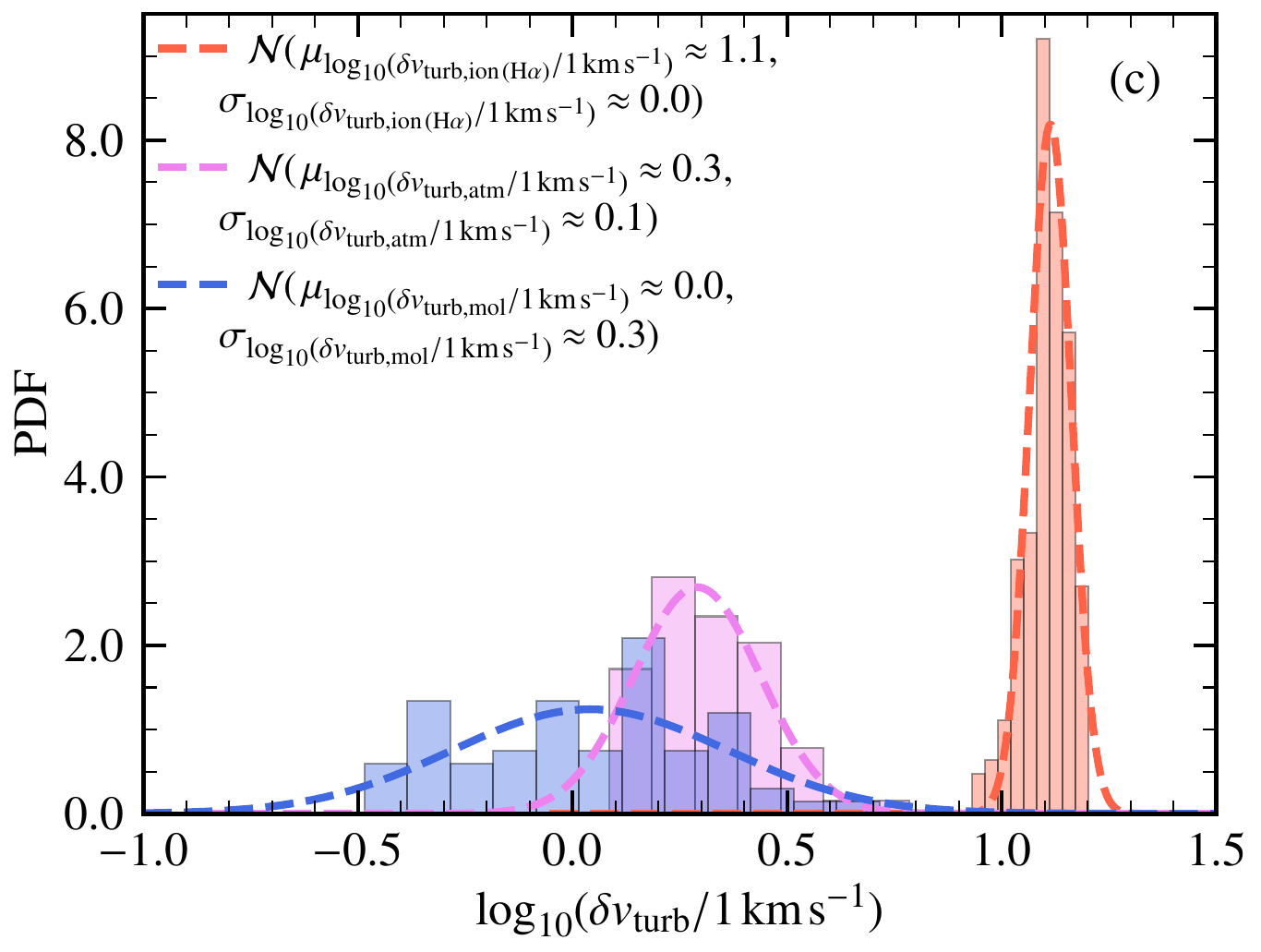}  \hspace{0.5cm}
\includegraphics[width=\columnwidth]{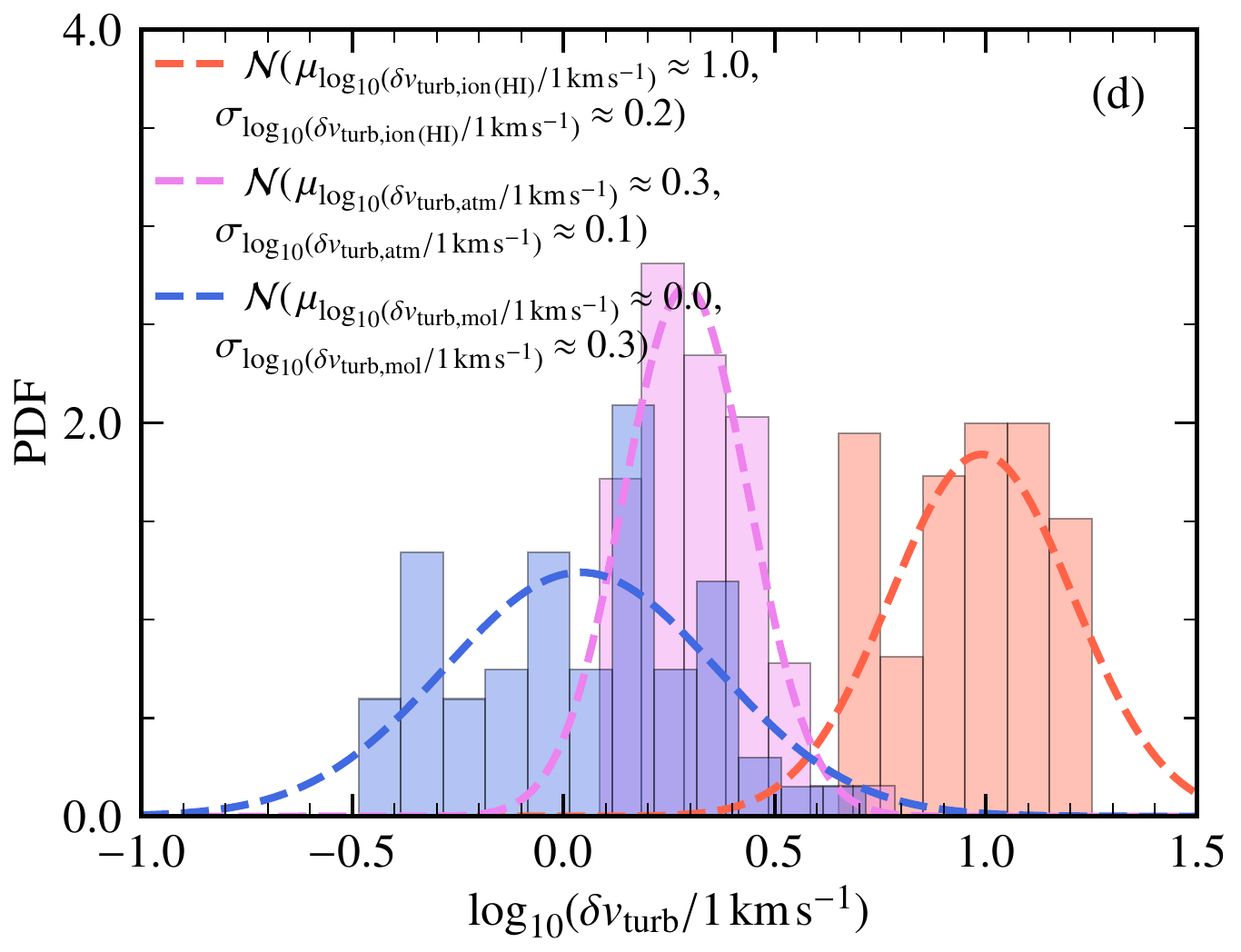}
\caption{
Probability density functions (PDFs) for gas density, $n$~(panel~(a)), the absolute value of line-of-sight magnetic fields, $|\Blos|$~(panel~(b)), and turbulent velocities, $\dvtHa$~(panel~(c)) and $\dvtHI$~(panel~(d)), for ionised (ion), atomic (atm), and molecular (mol) phases of the ISM. For all the distributions, we fit a lognormal function to the PDF \rev{(the current data size and quality is insufficient to fit further complicated models, see the text in \Sec{sec:pdf} for further discussion)}, and the estimated mean and standard deviation of the underlying normal distribution are reported in the legend. The sharp peak for $\dvtHa$ at $\log_{10}(\dvtionHa/1\,\km\,\s^{-1}) \approx 1.1$, which implies $\langle \dvtHa \rangle \approx 12.6\,\km\,\s^{-1}$, might be due to the large channel width of the H$\alpha$ spectra ($= 12\,\km\,\s^{-1}$). In comparision, $\dvtHI$ (spectral channel width $\approx 1\,\km\,\s^{-1}$) distribution shows a much wider distribution and might better probe the turbulence along the pulsar sightlines. Even though the densities between the three phases are well-separated, there is a significant overlap in the corresponding magnetic field distributions. The width of the magnetic field distribution is highest for the molecular ISM and that is a combination of (relatively wider) turbulent velocity and density distributions (note that the density distribution is the widest for the ionised ISM). This further emphasises that the properties of the ISM magnetic fields are determined by both the density and turbulent velocity and not just the density. The corresponding quantitatification in terms of the mean and standard deviation of the fitted distributions is given in \Tab{tab:stat}. 
}
\label{fig:pdf}
\end{figure*}

\Fig{fig:pdf} shows the probability density functions (PDFs) for $n, |\Blos|, \dvtHa,$ and $\dvtHI$ for all three ionised (probed by pulsars, H$\alpha$, and HI data), atomic (probed by $\ZHI$), and molecular (probed by $\ZMol$) ISM. We fit a normal function, $\mathcal{N}(\mu, \sigma)$, to the PDF of $\log_{10}$ of the variable for all the distributions assuming that the underlying variable follows a lognormal distribution. The fit is not equally good for all variables, especially $n$ and $|\Blos|$ in the molecular ISM show some sign of a bi-modal distribution. So, here, the normal function fit primarily captures the dominant peak. We discuss the choice of the lognormal functional form and, based on the fitted parameters (given in \Tab{tab:stat}), the phase-wise properties and implications of each distribution below. 

\begin{table}
\centering
\caption{The mean ($\mu$), standard deviation ($\sigma$), and relative level of fluctuations ($\sigma/\mu$) of density ($n$), absolute value of the line-of-sight magnetic field ($|\Blos|$), and turbulent velocity ($\dvt$) for each ISM phase, ionised (ion), atomic (atm), and molecular (mol), obtained from the fit results shown in \Fig{fig:pdf}.}
\label{tab:stat}
\begin{tabular}{ccccc} 
\hline
\hline
Quantity & Phase & $\mu$ & $\sigma$ & $\sigma/\mu$ \\
\hline
\hline
$n$ & ion & $\approx 0.025\,\cm^{-3}$ & $\approx 0.035\,\cm^{-3}$ & $\approx 1.4$ \\ 
$n$ & atm & $\approx 63\,\cm^{-3}$ & $\approx 30\,\cm^{-3}$  & $\approx 0.5$ \\ 
$n$ & mol & $\approx 3200\,\cm^{-3}$ & $\approx  2400\,\cm^{-3}$ & $\approx 0.7$ \\ 
\hline
$|\Blos|$ & ion & $\approx 1.3\,\muG$ & $\approx 0.9\,\muG$ & $\approx 0.7$ \\ 
$|\Blos|$ & atm & $\approx  4.0\,\muG$ & $\approx  3.0\,\muG$ & $\approx 0.7$ \\ 
$|\Blos|$ & mol & $\approx  6.3\,\muG$ & $\approx  15.0\,\muG$ & $\approx 2.4$ \\ 
\hline
$\dvtHa$ & ion & $\approx 12.6\,\km\,\s^{-1}$ & $\approx 0.0\,\km\,\s^{-1}$ & $\approx 0.0$ \\ 
$\dvtHI$ & ion & $\approx 10.0\,\km\,\s^{-1}$  & $\approx 4.8\,\km\,\s^{-1}$  & $\approx 0.5$ \\ 
$\dvt$ & atm & $\approx 2.0\,\km\,\s^{-1}$ & $\approx 0.4\,\km\,\s^{-1}$ & $\approx 0.2$ \\ 
$\dvt$ & mol & $\approx 1.0\,\km\,\s^{-1}$ & $\approx 0.7\,\km\,\s^{-1}$ & $\approx 0.7$ \\ 
\hline
\hline
\end{tabular}
\end{table}

\subsubsection{PDFs of $n$} \label{sec:pdfn}

A lognormal density distribution is usually associated with the turbulent origin of density structures in the ISM \citep{ElmegreenS2004}. In the cold, molecular ISM, the gas density distribution is usually assumed to follow a lognormal \citep{Vazquez-Semadeni1994, PassotV1998, FederrathEA2008} or Hopkins \citep[][]{Hopkins2013, FederrathB2015, SquireH2017, BeattieEA2022} PDF. The density PDF in star-forming regions might be even more complicated and, in addition to lognormal distribution, may include one or more power-law tails at higher densities \citep{Burkhart2018, BurkhartP2019, KhullarEA2021, AppelEA2022, AppelEA2023, MathewEA2024} \rev{or can be just a power-law \citep{AlvesEA2017}}. However, given the small sample size of the data in the $\ZMol$ dataset ($\nsamp=68$), it is difficult to fit these complicated models and we only chose to fit a lognormal distribution. In the atomic and ionised regions, the multiphase ISM simulations show lognormal density distribution \citep{deAvillezB2005, GentEA2013, SetaF2022}. The density distribution is also observationally shown to roughly follow a lognormal distribution in both the atomic \citep{BurkhartEA2015} and ionised \citep{BerkhuijsenF2008, BerkhuijsenF2015} ISM.

The density distribution for all three phases and their corresponding lognormal fits are shown in \Fig{fig:pdf}~(a).  As expected, the mean of the distribution is significantly different between the phases, the mean of $n$ is $\approx 0.025\,\cm^{-3}$ in the ionised ISM, $\approx 63\,\cm^{-3}$ in the atomic ISM, and $\approx 3200\,\cm^{-3}$ in the molecular ISM. However, the relative density distribution is the widest in the diffuse ($\approx 1.4$) ionised ISM and has more comparable widths in the atomic ($\approx 0.5$) and molecular ($\approx 0.7$) phases (see \Tab{tab:stat}).

\subsubsection{PDFs of $\Blos$} \label{sec:pdfblos}

The structure of magnetic fields in the ISM is determined by a variety of physical processes: tangling of the large-scale (galactic, $\kpc$ scale) field, turbulent dynamo on scales comparable to the driving scale of ISM turbulence ($\approx 100\,\pc$ for supernova explosions), and compression due to shocks (at a variety of scales). The turbulent dynamo and shock compression produce strongly intermittent, non-Gaussian magnetic fields  \citep{ZeldovichEA1984, RuzmaikinEA1989, SchekochihinEA2004, SetaEA2020, SetaF2021dyn, SetaF2022, SurS2024}, whereas the tangling of the large-scale field will likely produce Gaussian fields due to the volume filling nature of the ISM turbulence (see Appendix~A in \citealt{SetaEA2018} and Sec.~4.1 in \citealt{SetaF2020}). Thus, the resulting magnetic field would be a mixture of both Gaussian and non-Gaussian components. Given that the contribution from each component is not known and the small sample size ($\nsamp = 212, 66,$ and $68$ in ion, atm, and mol phases), we assume $\Blos$ follows a normal distribution and thus fit a lognormal function to the PDF of $\log_{10}(|\Blos|/1\,\muG)$.

\Fig{fig:pdf}~(b) shows the PDF of $\log_{10}(|\Blos|/1\,\muG)$ for all three ion, atm, and mol phases. The mean value of magnetic fields is stronger in the denser regions, increasing from the ionised ($\approx 1.3\,\muG$) to atomic ($\approx 4.0\,\muG$) to molecular ($\approx 6.3\,\muG$) phases but there is a significant overlap between the three PDFs. Moreover, the width in the ionised ISM ($\approx 0.9\,\muG$) is lower than that in the cold atomic phase ($\approx 3\,\muG$) but the width is significantly higher in the molecular phase ($\approx 15\,\muG$). This further demonstrates that the magnetic fields respond to the ISM properties other than the gas density (compare relative widths in \Fig{fig:pdf}~(a) and \Fig{fig:pdf}~(b), also see \Tab{tab:stat}) and the $B \propto n^{\rm factor}$ does not hold across all the ISM phases as $\Emag \propto \Ekin$ does (\Fig{fig:emagekin}). Moreover, the relative fluctuations (standard deviation/mean) in the ionised and atomic phases are similar ($\approx 0.7$), it is significantly higher in the molecular phase ($\approx 2.4$).

\subsubsection{PDFs of $\dvtHa$ and $\dvtHI$} \label{sec:pdfdvturb}

Turbulence in the ISM is driven by a variety of processes at a range of scales \citep{ElmegreenS2004, ScaloE2004, MacLowK2004} and the ISM turbulence is also expected to be a combination of spatially intermittent, non-Gaussian and Gaussian components \citep[see Sec.~3.2 in][for a discussion]{HennebelleF2012}. However, like magnetic fields, due to the lack of proper characterisation of the ISM turbulent velocities and given the limited data size, we assume that the underlying turbulent velocities follow a normal distribution and fit a lognormal function to the PDF of $\log_{10}(\dvt/1\,\km\,\s^{-1})$.

\Fig{fig:pdf}~(c) and \Fig{fig:pdf}~(d) shows the PDFs of  $\log_{10}(\dvt/1\,\km\,\s^{-1})$ and their corresponding fits for all three phases. It is inherently difficult to directly compare $\dvtHa$ and $\dvtHI$ distributions due to an order of magnitude difference in their spectral velocity channel widths ($\approx 1\,\km\,\s^{-1}$ for HI and $12\,\km\,\s^{-1}$ for H$\alpha$). For the $\dvtHa$ case (\Fig{fig:pdf}~(c)), there is practically no overlap between the turbulent velocities in the ionised phase and the other two phases, which show some overlap. However, we caution that the mean of $\dvtHa$ distribution $\approx 12.6\,\km\,\s^{-1}$ (very close to the channel width) and very small width might be due to the large-channel width of the H$\alpha$ spectra. For the $\dvtHI$ case (\Fig{fig:pdf}~(d)), the mean value of turbulent velocity distribution decreases from the ionised ($\approx 10\,\km\,\s^{-1}$) to atomic ($\approx 2\,\km\,\s^{-1}$) to molecular ($\approx 1\,\km\,\s^{-1}$) ISM but the relative width is highest for the molecular phase ($0.7$, see \Tab{tab:stat}). So, even though the density shows a lower relative width in for the molecular ISM (\Fig{fig:pdf}~(a)), the large width in the magnetic field distribution (\Fig{fig:pdf}~(b)) might be due to a wider turbulent velocity distribution. Moreover, for the ionised ISM, the wider density distribution does not necessarily result in a wider magnetic field distribution probably because of comparatively less wide turbulent velocity distribution (compare numbers across phases in \Tab{tab:stat}). These inferences further emphasise that the ISM magnetic fields are determined by both the density and turbulent velocities and not just the density.

\begin{table}
\centering
\caption{The density ($n$), temperature ($T$),  absolute value of the line-of-sight magnetic field ($|\Blos|$), and plasma beta ($\pbeta$, \Eq{eq:beta}, \rev{the upper limit of the conventional plasma beta since only the line-of-sight field strength used}) for all three phases. $n$ and $|\Blos|$ are taken from \Tab{tab:stat} and $T$ is assumed for each phase to compute $\beta_{\rm plasma}$. In all the phases, the magnetic pressure is comparable to the thermal pressure, further demonstrating that magnetic fields are an important component of all the ISM phases.}
\label{tab:beta}
\begin{tabular}{ccccc} 
\hline
\hline
Phase & $n\,[\cm^{-3}]$ & $T\,[\K]$ & $|\Blos|\,[\muG]$ & $\beta_{\rm plasma}$\\
\hline
\hline
Ionised & $\approx 0.025\,\cm^{-3}$  & $\approx 10^{4}\,\K$ & $\approx 1.3\,\muG$ & $\approx 0.5$ \\ 
Atomic & $\approx 63\,\cm^{-3}$ & $\approx 10^{2}\,\K$ & $\approx  4.0\,\muG$ & $\approx 1.4$ \\ 
Molecular & $\approx 3200\,\cm^{-3}$ & $\approx 10\,\K$ & $\approx  6.3\,\muG$ & $\approx 3.0$ \\ 
\hline
\hline
\end{tabular}
\end{table}

\subsection{Phase-wise importance of magnetic fields} \label{sec:beta}

The importance of magnetic fields is usually characterised by plasma beta, which is the ratio of the thermal to magnetic pressure. We define it as
\begin{align} \label{eq:beta}
\pbeta = n k_{\rm B} T / (\Bloss /8 \pi),
\end{align}
where $n$ is the number density, $k_{\rm B}$ is the Boltzmann constant, $T$ is the temperature, and $\Blos$ is the line-of-sight magnetic field. In principle, the definition involves the total magnetic field strength but we use the available $\Blos$ values, so, $\pbeta$ computed this way is \rev{actually the upper limit of the plasma beta}. The higher values of $\pbeta$ ($\gg 1$) imply that magnetic fields have negligible impact and could be ignored, $\pbeta \approx 1$ correspond to magnetic pressure being as important as the thermal pressure, and $\pbeta \ll 1$ corresponds to the situation where magnetic pressure is dynamically more important than the thermal pressure. From this work, we have phase-wise information of $n$ and $\Blos$, and we source $T$ in each phase from the literature. We assume $T\approx10^{4}\,\K$ for the ionised phase (warm ionised gas probed by the pulsars), $T\approx 10^{2}\,\K$ for the atomic phase (the dense gas, see \Tab{tab:stat}, probed by the Zeeman effect is majorly the cold atomic medium), and $T\approx10\,\K$ for the molecular phase \citep{Ferriere2001, Ferriere2020}. The estimated $\pbeta$ values are given in \Tab{tab:beta}. \rev{We emphasise that these values represent typical numbers in each phase and, in principle, the plasma beta would also have a distribution.} For all the phases, we find $\pbeta \approx 1$, and thus the magnetic pressure is comparable to the thermal pressure. This further demonstrates that the magnetic fields are an important component of all the ISM phases and their dynamical impact should not be ignored.

\section{Discussion} \label{sec:dis}

\subsection{Further confirmation of the analysis for the ionised ISM} \label{sec:pul}
For the results and discussion in \Sec{sec:res}, we treated all three observational datasets, pulsars, $\ZHI$, and $\ZMol$ equally. However, there are two primary differences between the analysis with the pulsar data and Zeeman measurements. First, the turbulent velocities for the Zeeman measurements are derived using the Zeeman data itself, whereas, for pulsars, the turbulent velocities are estimated from independent H$\alpha$ and HI observations. Second, and more importantly, the gas density and line-of-sight magnetic fields in the Zeeman data probe localised regions of the ISM, whereas the pulsars give the {\it average} gas density (\Eq{eq:meanne}) and line-of-sight magnetic fields (\Eq{eq:avgbpar}) over the entire path-length. We assume these average values as representative numbers in the ionised ISM. Here, using the observational dataset (\Sec{sec:len}) and multiple multiphase ISM simulations (\Sec{sec:sim}), we demonstrate that these are fair assumptions and further confirm our analysis using the pulsar and HI data for the ionised ISM.

\subsubsection{Length scales in the ionised ISM} \label{sec:len}

\begin{figure*}
\includegraphics[width=\columnwidth]{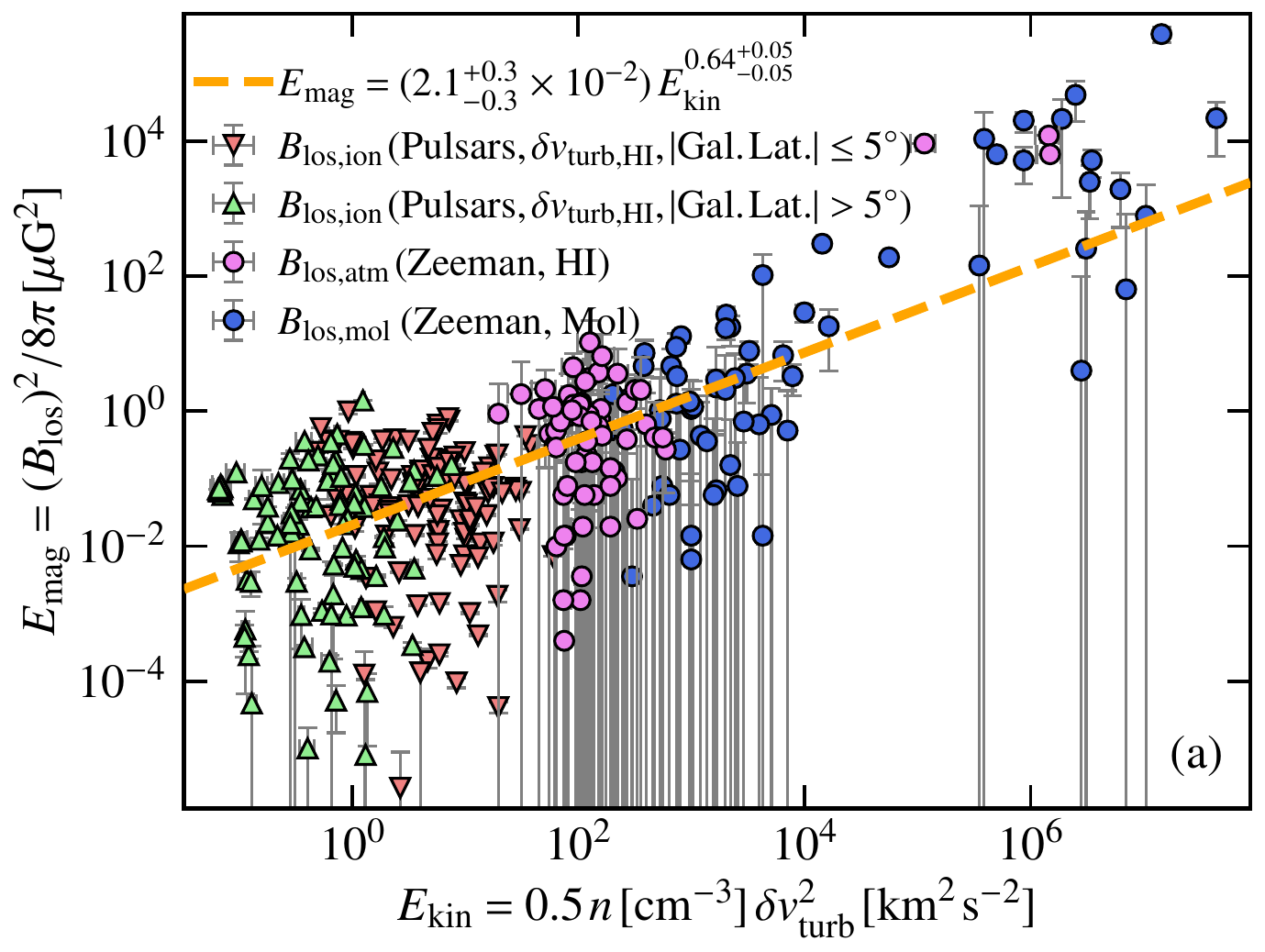}  \hspace{0.5cm}
\includegraphics[width=\columnwidth]{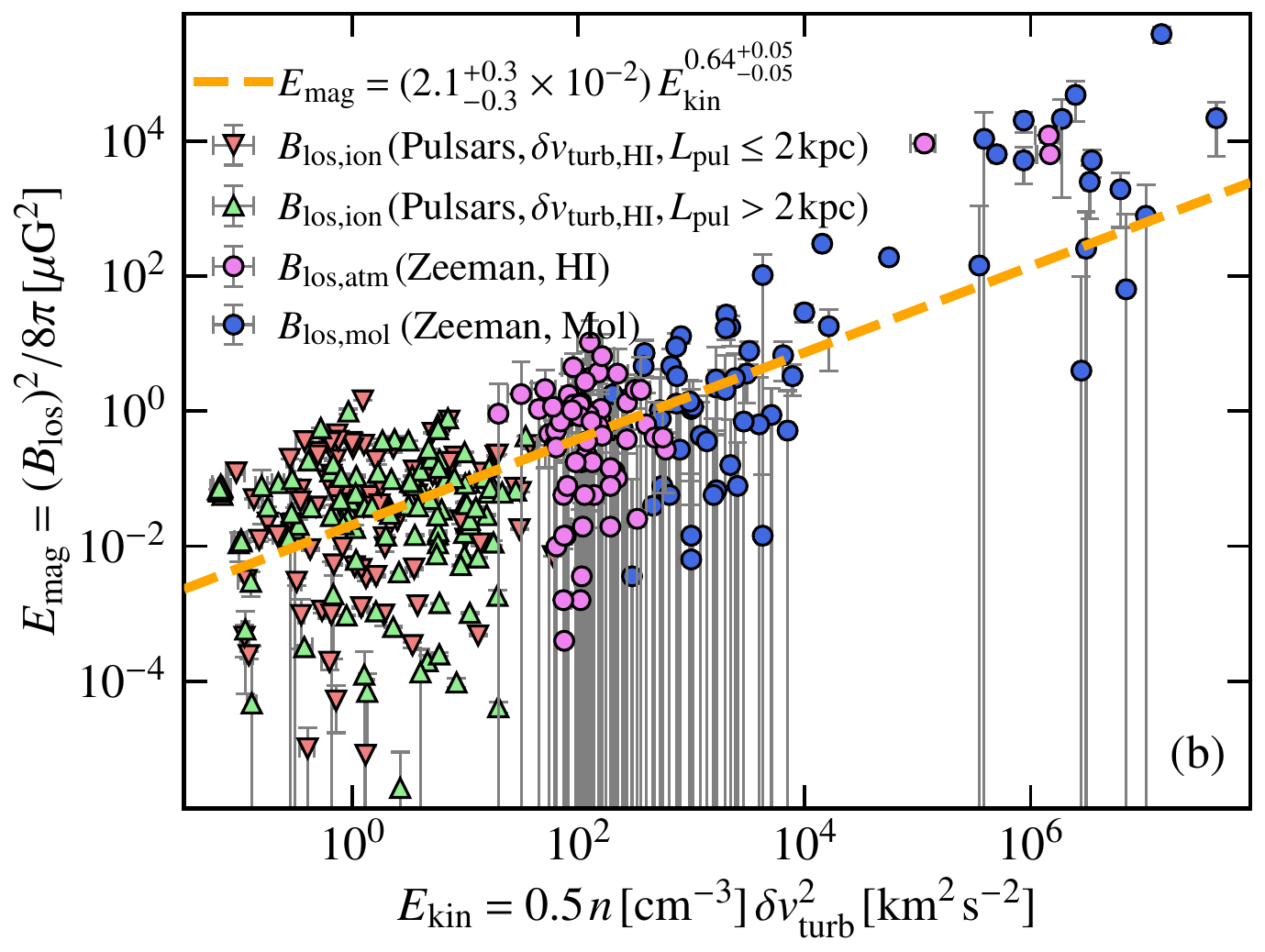}
\caption{To test the influence of averaging over large, of the order of $\kpc$s, length scales to derive the $\Emag$ -- $\Ekin$ relation in the ionised ISM from the pulsar observations (\Sec{sec:pdata}), we divide the data into groups by the Galactic latitude (Gal.~Lat.) of the pulsar (panel~(a)) and the distance to the pulsar (panel~(b), $\Lpul$). When divided by the Gal. Lat.~(a), Gal.~Lat.$\le 5^{\circ}$ ($\nsamp=108$) and Gal.~Lat.$> 5^{\circ}$ ($\nsamp=104$), the pulsars in the lower Galactic latitude tend to probe larger values of  $\Ekin$ due to higher densities and turbulent velocities in the Galactic plane. But neither group shows a preference to be closer to $\Emag \propto \Ekin$ fit (dashed orange line). On the other hand, when divided by $\Lpul$, $\Lpul \le 2\,\kpc$ ($\nsamp=85$) and $\Lpul > 2\,\kpc$ ($\nsamp=127$), the pulsars are randomly distributed along and across the $\Emag \propto \Ekin$ fit. These tests confirm that the average values of density and line-of-sight magnetic fields used to compute $\Emag$ and $\Ekin$ from the pulsar sample are representative of the ionised ISM inspite of the averaging.
}
\label{fig:emagekinpul}
\end{figure*}

To test the effect of averaging over large length scales ($=\Lpul$, distance to the pulsar) in the analysis for the ionised ISM, we divide the pulsar sample into two groups. \rev{Once by the Galactic latitude, $|\GL|=5^{\circ}$ (\Fig{fig:emagekinpul}~(a)), and next by the distance to the pulsar, $\Lpul=2\,\kpc$ (\Fig{fig:emagekinpul}~(b)). For both cases, the motivation is to test whether the $\Emag \propto \Ekin$ relation shows any variation with more of the ISM along the line-of-sight -- either by probing more of the Galactic plane or due to a longer line of sight. The cutoff Galactic latitude, $|\GL|=5^{\circ}$, and distance, $\Lpul=2\,\kpc$,  are reasonably chosen based on the fact that they yield roughly similar numbers of pulsars in each latitude/distance bin.} 

In \Fig{fig:emagekinpul}~(a), we separate the pulsars using the Galactic latitude ($\GL$) to probe the ISM close to the Galactic plane ($|\GL| \le 5^{\circ}$) and the pulsars located away from the plane ($|\GL| > 5^{\circ}$). The pulsars close to the plane probe the ionised ISM with statistically larger values of $\Ekin$ (due to higher $n$ and $\dvt$), while that for away samples are averaged over regions with significantly lower values of $n$ and $\dvt$. This gives rise to some separation along the $\Emag \propto \Ekin$ fit (dashed orange line in \Fig{fig:emagekinpul}~(a)) but there is no preference for either group to be close to the fitted line. Both groups show an equal level of spread across it.

Next, in \Fig{fig:emagekinpul}~(b), we divide the pulsar sample into two groups based on the distance to the pulsar, $\Lpul \le 2\,\kpc$ and $\Lpul > 2\,\kpc$. Both groups show equal spread along and across the $\Emag \propto \Ekin$ fit (dashed orange line in \Fig{fig:emagekinpul}~(b)). These tests (\Fig{fig:emagekinpul}) demonstrate that averages of thermal electron density and line-of-sight magnetic fields over the $\kpc$ distances to the pulsars are good representatives of typical density and line-of-sight magnetic field strength in the ionised ISM.

\subsubsection{Confirmation from multiphase ISM simulations} \label{sec:sim}

To further check the $\Emag \propto \Ekin$ relation in the ionised ISM phase, we use two types of multiphase ISM simulations: numerically driven turbulence, two-phase simulations from \citet{SetaF2022} and supernova driven, three-phase $\TIGRESS$ (Three-phase Interstellar Medium in Galaxies Resolving Evolution with Star Formation and Supernova Feedback) simulations from \citet{Kado-FongEA2020}. We first give here basic overall details of each type of simulation and then describe the construction of $\Emag$ and $\Ekin$ using it.

The first type of simulation is taken from \citet{SetaF2022}, where turbulence is driven numerically at a scale of $100\,\pc$ in a three-dimensional, triply-periodic domain ($512^{3}$ grid points) of size $200\,\pc$ along each dimension with Milky Way-type heating and cooling functions. The turbulence is driven both solenoidally, $\Sol$, and compressively, $\Comp$ \citep[see][for further details]{FederrathEA2008}. The equations for non-ideal magnetohydrodynamics are solved and the corresponding viscosity and resistivity are chosen to be a factor of $\approx 4$ times higher than their expected numerical values. Further details of the setup and initial conditions are described in Sec.~2 of \citet{SetaF2022}. The initial weak magnetic fields (random field of strength $10^{-4}\,\muG$) amplify due to the turbulent dynamo mechanism, achieving a statistically steady state. We take hydrogen number density ($n_{\rm H}$), temperature ($T$), velocity, and magnetic fields from a snapshot in this statistically steady state. The thermal electron density, $\ne$, needed to compute mock $\DM$ and $\RM$ from these simulations is computed using $n_{\rm H}$ and $T$ as \citep{HollinsEA2017}
\begin{equation} \label{eq:nesim}
\ne = n_{\rm H} \left[\frac{\arctan(T/10^{3}\,\K - 10)}{\pi} + \frac{1}{2}\right].
\end{equation} 

For the second type, we take $\TIGRESS$ simulations (data sourced from \href{https://princetonuniversity.github.io/astro-tigress}{https://princetonuniversity.github.io/astro-tigress}), which represents the solar neighbourhood conditions and includes galactic shear, self-gravity, star particles, turbulence driven by self-consistent supernova feedback, and relevant cooling and heating functions \citep{KimO2017}. They start their simulations with a relatively strong magnetic field (uniform field of strength $2.6\,\muG$) and solve ideal magnetohydrodynamics equations for $660\,\Myr$. The chosen model (\texttt{R8\_2pc.0300}) has a size of $1024\,\pc \times 1024\,\pc \times 7168\,\pc$ ($xy$ represent the Galactic plane and $z$ the vertical direction) with a uniform grid spacing of $2\,\pc$ and is at a time of $300\,\Myr$ in the evolution. Then, using $n_{\rm H}$, $T$, and properties of star particles, the authors post-process the data using ray-tracing and ionisation calculation \citep[see Sec.~2 in][for further details]{Kado-FongEA2020} to determine the equilibrium fraction of the atomic component, which can be further used to compute $\ne$. From the $\TIGRESS$ dataset, we take $n_{\rm H}$, $T$, velocity, magnetic fields, and $\ne$.

\begin{figure}
\includegraphics[width=\columnwidth]{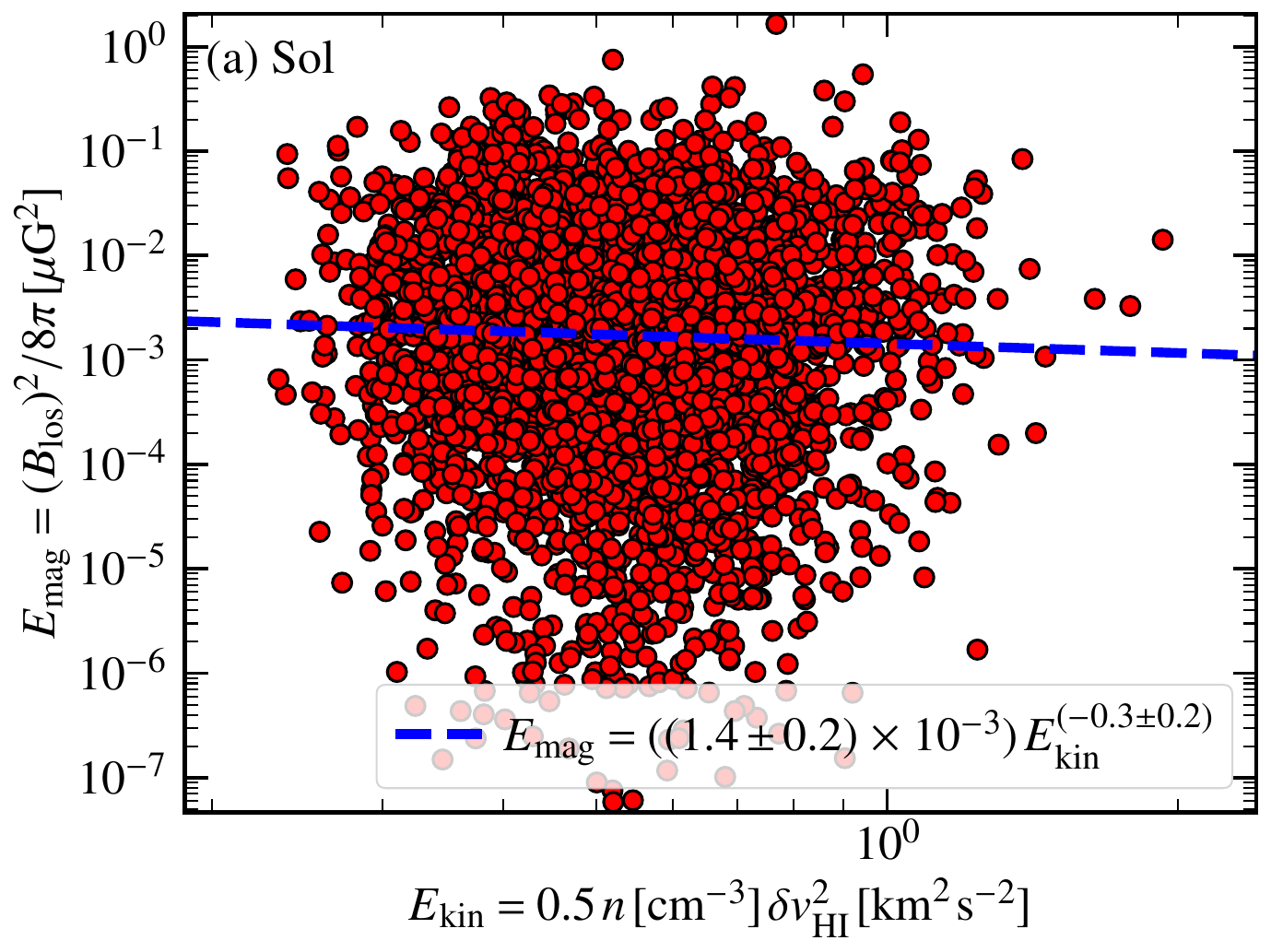} \\
\includegraphics[width=\columnwidth]{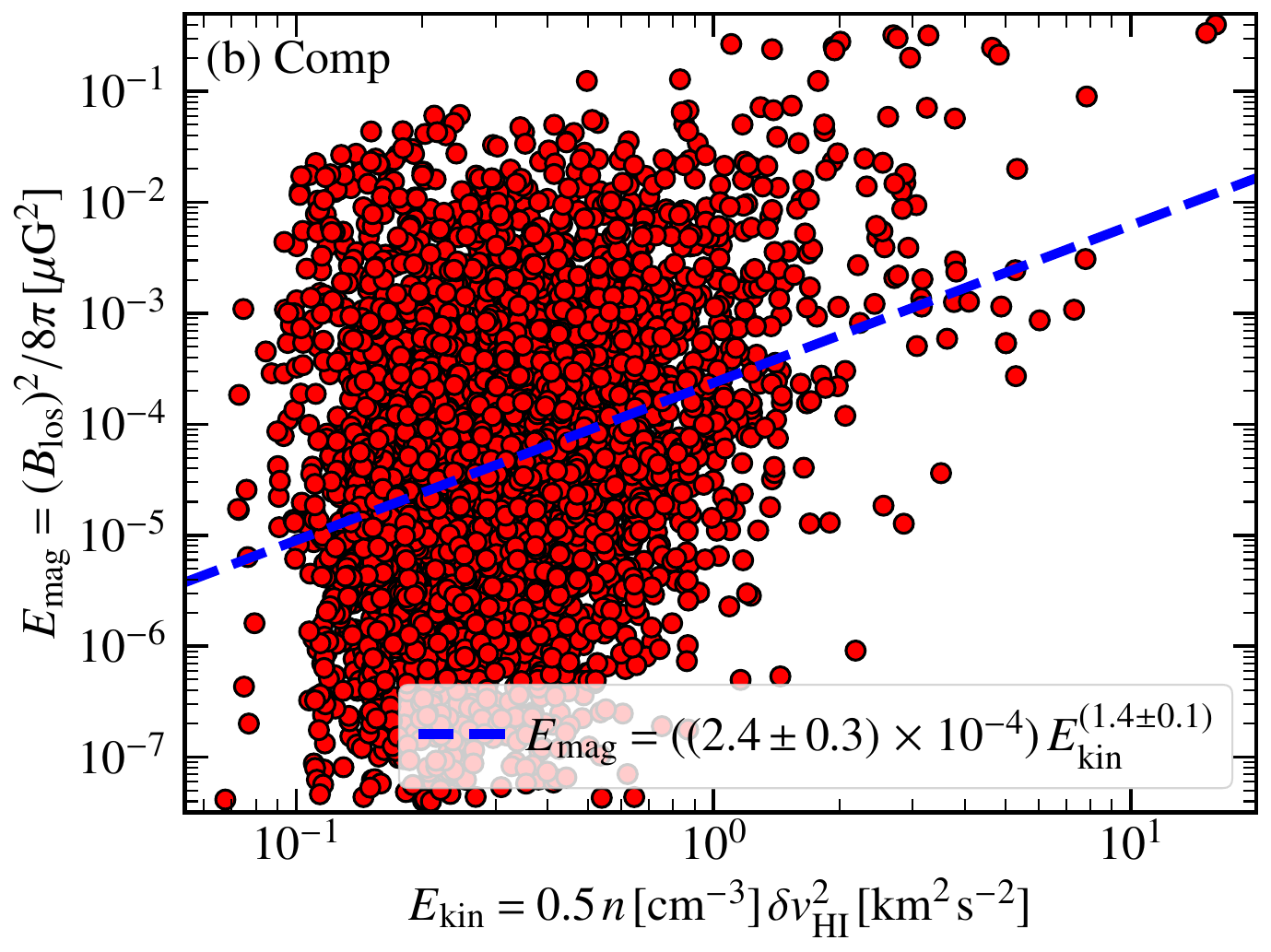} \\
\includegraphics[width=\columnwidth]{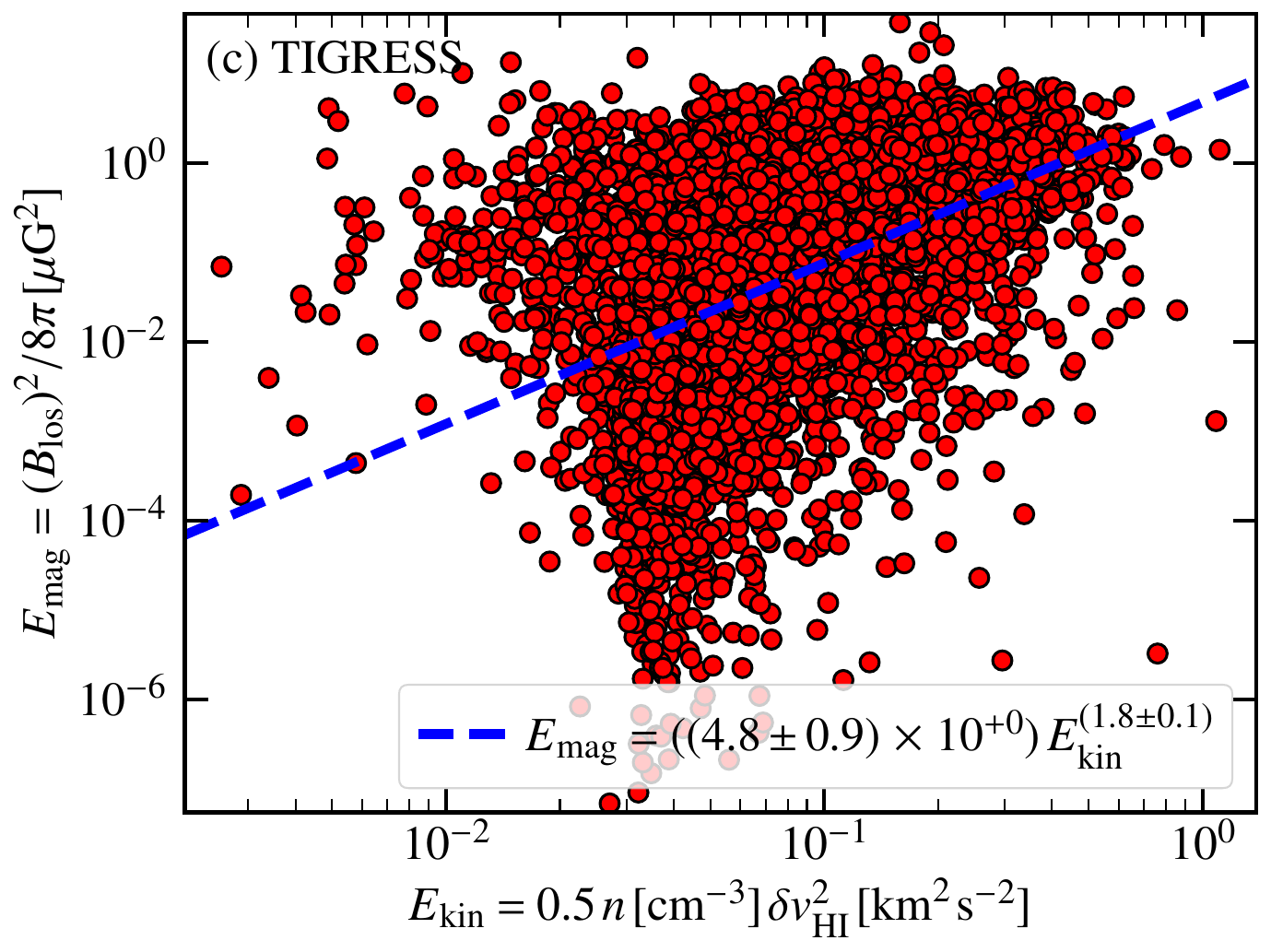}
\caption{The $\Emag$ -- $\Ekin$ relation derived from observations (\Fig{fig:emagekin}) is tested only for the ionised ISM using two-types of multiphase ISM simulations: two-phase, numerically driven turbulence simulations ($\Sol$, panel~(a) and $\Comp$, panel~(b)) and three-phase, supernova driven turbulence simulations ($\TIGRESS$, panel~(c)). For these simulations, we use the analysis exactly as done for the ionised ISM (\Sec{sec:pdata} and \Sec{sec:vdata}). From the mock observations, for each case, 10000 points (red) are randomly selected and fitted with a power-law (dashed blue line). For the $\Sol$ case, $\Emag$ does not show a significant dependence on $\Ekin$ and this is due to a very small range of values in $\Ekin$ ($x$-range). On the other hand, both $\Comp$ and $\TIGRESS$ cases show a power-law dependence.  The exact coefficient and exponent in the power-law fit depends on the physics included in the simulations but $\Emag \propto \Ekin$ holds true for both those cases.
}
\label{fig:emagekinsim}
\end{figure}

We use the data from these simulations and repeat the entire analysis done for the ionised ISM using pulsar observations (\Sec{sec:pdata} and \Sec{sec:vdata}). First, knowing $\ne$, choosing a line-of-sight (along $\hat{y}$ for both types of simulations), and utilising the entire domain as path length ($200\,\pc$ for $\Sol$ and $\Comp$ datasets and $1024\,\pc$ for the $\TIGRESS$ data), we compute mock $\DM$ and $\RM$ using \Eq{eq:dm} and \Eq{eq:rm}, respectively. Then, from the mock $\DM$ and $\RM$, using \Eq{eq:meanne} and \Eq{eq:avgbpar}, we compute $\langle \ne \rangle$ and $\langle \Blos \rangle$. As with the pulsar data, we assume that these represent typical gas density and line-of-sight magnetic field for the ionised ISM, i.e.,~$n \approx \langle \ne \rangle$ and $\Blos \approx \langle \Blos \rangle$. Next, using density, temperature and line-of-sight velocity (velocity along $\hat{y}$ direction), we perform the radiative transfer calculation to generate mock HI emission spectra using the formalism in Sec~2.2 of \citet{BhattacharjeeEA2024} (also, see \citealt{Miville-DeschenesEA2003}). Then, using \Eq{eq:m2}, we compute $\dv$ for each line-of-sight and, like for pulsars, assume the Mach number in the ionised ISM $\Mach_{\rm ion} \approx 1$ to compute $\dvtHI$. Finally, using $\Blos$, $n$, and $\dvtHI$, $\Emag$ and $\Ekin$ are constructed. 

From the obtained $\Emag$ and $\Ekin$, $10000$ random points are selected and are fit with a power-law, $\Emag = \mathcal{C}_{E}\,\Ekin^{\alpha_{E}}$, for all three simulated datasets, $\Sol$, $\Comp$, and $\TIGRESS$. The data and the determined fit for each simulated dataset are shown in \Fig{fig:emagekinsim}. The $\Sol$ case does not show any significant relation between $\Emag$ and $\Ekin$, but this might also be because of a small range of $\Ekin$ ($x$-axis in \Fig{fig:emagekinsim}~(a)). \rev{The large scatter seen in the simulated data compared to the observations in \Fig{fig:emagekin} may be more evident here due to the limited range on the $x$-axis or is expected at lower densities because of the physics of different magnetohydrodynamic modes \citep{PassotV2003, Vazquez-SemadeniEA2024}.} For cases with a significant $\Ekin$ range (at least an order of magnitude), $\Comp$ in \Fig{fig:emagekinsim}~(b) and $\TIGRESS$ in \Fig{fig:emagekinsim}~(c), $\Emag \propto \Ekin$. Both the coefficient, $\mathcal{C}_{E}$, and exponent, $\alpha_{E}$, are significantly different between the two and those in the observational results shown in \Fig{fig:emagekin}. This might be due to differences in the physics modelled in these simulations and further complexities in observations (line-of-sight effects, the influence of the Local Bubble, bias in pulsar locations, multiscale magnetic field and thermal electron density structures, observational effects, etc.). Understanding these differences requires further work and more pulsars with independent distances. However, the general trend of proportionality between the $\Emag$ and $\Ekin$ confirms our method for the ionised ISM, especially the idea of combining $n$ obtained from the pulsar $\DM$ observations with $\dvt$ from HI spectra to compute $\Ekin$.
  
\subsection{ISM magnetic fields are determined by both the density and turbulent velocity} \label{sec:comb}
In \Sec{sec:res}, using both the $\Emag$ -- $\Ekin$ relation (\Fig{fig:emagekin}) and relative differences between the PDFs of density, magnetic fields, and turbulent velocities with the ISM phase (\Fig{fig:pdf} and \Tab{tab:stat}), we demonstrated that the magnetic field depends on both the density and turbulent velocity and not just the density as, at times, inferred from the magnetic field - density relations. However, we showed this only for the magnetic field strengths. The magnetic field structure might also depend on both the density and turbulent velocity structures. In the literature, both aspects are individually discussed. The atomic hydrogen filaments are shown to be aligned with the ISM magnetic fields, suggesting a more density-magnetic field connection in terms of the structure \citep{ClarkEA2014, ClarkEA2019, MaEA2023}. There are also discussions associating these structures more with the turbulent velocity properties \citep{LazarianP2000, YuenEA2019, YuenEA2021}. In principle, both could be right and the contribution from each would depend on the phase, length scales, and the type of observable. For example, the synchrotron polarisation probes more of the warm, ionised ISM, where turbulent velocities might play a dominant role and the dust polarisation probes more of the colder ISM, where the density fluctuations might play a major role. Separating the contribution from the density and turbulent velocity structures to magnetic structures requires further work, which we aim to do in the future.

\subsection{Assumptions and missing effects} \label{sec:miss}

Throughout the analysis, we have made some simplifying assumptions and, here, we discuss them and their possible implications on the derived results. 

To compute the velocity widths from the all-sky H$\alpha$ and HI spectra, we associated features within velocities of $\pm\,50\,\km\,\s^{-1}$ with the Milky Way. This choice is motivated by the fact we would like to avoid the isolated high-velocity clouds \citep[usually at absolute velocities greater than $90\,\km\,\s^{-1}$, see][]{Wakkerv1997} as they may not represent the typical ISM along pulsar sight lines, which samples a very heterogeneous region of the Milky Way's ISM. Moreover, from a larger HI absorption sample \citep[\texttt{BIGHICAT} from][]{McClure-GriffithsEA2023}, most Milky Way features are observed within absolute velocities of $\approx \pm\,60\,\km\,\s^{-1}$ \citep{RybarczykEA2024}. We vary our limits of cutoff velocities from $\pm\,50\,\km\,\s^{-1}$ to $\pm\,60\,\km\,\s^{-1}$. This made negligible difference to results in \Fig{fig:emagekin} and \Fig{fig:pdf}~(c, d) and thus do not alter any of our conclusions.

For the ionised ISM, when computing $\dvt$ for $\Ekin$ from the HI emission spectra (\Sec{sec:vdata} and \Sec{sec:dvt}), we assumed the $\Mach_{\rm atm} \approx 1$. In principle, the HI emission spectra have contributions from both the warm and cold atomic, neutral medium \citep{WolfireEA1995, WolfireEA2003} and the cold atomic medium has significantly higher Mach number \citep[$\gtrsim 3$,][]{HeilesT2003II, MurrayEA2015, McClure-GriffithsEA2023}. However, \rev{on an average}, the HI emission spectra are dominated by the volume-filling, warm atomic component, so the contribution from the localised cold atomic medium (which is primarily probed by the HI absorption spectra) is \rev{sub-dominant}. Moreover, even though pulsar observations probe the ionised ISM, we associate the turbulent velocities computed from the HI spectra (in emission) with those. This assumes that, on average, the turbulent velocities in the atomic ISM are comparable to the turbulent velocities in the ionised ISM \citep[see Table~2 in][]{Ferriere2020}. Note that we also use the H$\alpha$ spectra, which indeed probes the ionised ISM and the $\Emag$ -- $\Ekin$ relation are very similar for both the H$\alpha$ and HI cases (\Fig{fig:emagekin}). \rev{Moreover, in principle, $\Mach$ in each phase might also have some variation with the line-of-sight but the currently available data is insufficient to include such a variation in the analysis.}

We have only considered the line-of-sight components of magnetic fields and turbulent velocities to compute $\Emag$ and $\Ekin$. In principle, the component perpendicular to the line of sight can be different for both and need not necessarily be correlated. This assumption also, in some sense, is connected to the isotropy of 3D small-scale structures in the ISM. These 3D structures might be anisotropic (e.g.~ filaments and sheets), and that might introduce corrections to the derived $\Emag \propto \Ekin$ relation. For the ionised ISM, the large distances to pulsars lead to density and magnetic fields being averaged over multiple such smaller-scale structures and this has been further tested in \Sec{sec:pul}. \rev{Moreover, for the ionised ISM, the angular beam size is significantly smaller for the pulsar data (pencil beam) in comparision to the H$\alpha$ ($1\,\deg$) and HI ($10.8$ arcmin or $16.2$ arcmin, depending on the survey, see \Sec{sec:vdata}) data. But we do not expect this to make a significant difference to the derived results as, again assuming isotropy of 3D small-scale structures in the ISM, averaging along the line-of-sight (for pulsar $\RM$s and $\DM$s) will have a similar effect as averaging on the plane-of-sky (for H$\alpha$ and HI spectra).}

We have assumed a single power-law function to capture the relationship between $\Emag$ and $\Ekin$ across all the ISM phases. From the multiphase simulations of the turbulent dynamo in the ISM, it is expected that the ratio of $\Emag/\Ekin$ depends on the phase of the ISM \citep[$\Emag/\Ekin \approx 0.1$ in the warm phase and $\approx 0.01$ in the cold phase, see][]{SetaF2022}. So, with the changing ISM phase, from ionised to atomic to molecular, the amplitude of the power-law function might change. This is difficult to explore with the current number of data points in each phase ($\nsamp=212$ in the ionised ISM, $\nsamp = 66$ in the atomic ISM, and $\nsamp=68$ in the molecular ISM) and given the large uncertainty in the magnetic field estimates. With much larger and more precise data, a more complicated fit function, possibly three power-laws with the same exponent but different coefficients, can be attempted.

Overall, with the available data, the analysis here allowed us to explore the relationship between $\Emag$ and $\Ekin$ over the three ISM phases (ionised, atomic, and molecular). In our future work, we aim to explore some of these additional effects using new observations and simulations.

\section{Summary and Conclusion} \label{sec:con}
Magnetic fields are an important component of the interstellar medium (ISM) of galaxies, but their connection with the multiphase ISM gas is yet to be completely known. Usually, to account for magnetic fields' impact on star formation, a relationship between the magnetic field strength, $B$, and gas density, $n$, is assumed. Here, for the Milky Way, in addition to Zeeman measurements of line-of-sight magnetic fields, $\Blos$, in the atomic and molecular ISM, we supplement $\Blos$ measurements in the ionised ISM using pulsar observations. This allows us to explore the magnetic fields across all three ISM phases, ionised, atomic, and molecular and by extension a large range of gas densities ($10^{-3}\,\cm^{-3} \lesssim n \lesssim 10^{7}\,\cm^{-3}$). In particular, we study the relationship between the turbulent kinetic energy, $\Ekin$, and magnetic energy, $\Emag$.

The Zeeman observations (described in \Sec{sec:zdata}) provide $n$, $\Blos$, and velocity widths of the components, $\dv$, but the pulsars observations only provide the average thermal electron density, $\langle \ne \rangle$ (\Eq{eq:meanne}), and average line-of-sight magnetic field, $\langle \Blos \rangle$ (\Eq{eq:avgbpar}), where averages are over the distance to the pulsar. So, for the ionised ISM, we perform two further steps in our analysis (\Sec{sec:pdata}). First, for each pulsar line-of-sight, we use both the ionised (H$\alpha$) and atomic (HI) hydrogen spectra to compute $\dv$ along that line of sight (\Sec{sec:vdata}). Second, we assume that, for the ionised ISM, gas with density $n \approx \langle \ne \rangle$ hosts line-of-sight magnetic fields $\Blos \approx \langle \Blos \rangle$ (tested and discussed using observations and multiphase simulations in \Sec{sec:pul}). Then, for all the three ISM phases, we assume a Mach number, $\Mach$, to compute the turbulent velocity, $\dvt$, from the observed (atomic and molecular) and computed (ionised) $\dv$, where we assume $\Mach_{\rm ion} \approx \Mach_{\rm atm} \approx 1$ and $\Mach_{\rm mol} \approx 5$ (\Sec{sec:dvt}). Finally, with the data ($n, \Blos, \dvt$) we compute the $\Emag = \Bloss/8\pi$ and $\Ekin=0.5\,n\,\dvtt$ for all three ionised, atomic, and molecular phases.

We find that $\Emag \propto \Ekin$ across all the ISM phases (\Fig{fig:emagekin} and \Sec{sec:mres}), and it is a more concrete relationship in comparision to $B \propto n^{\rm factor}$ as a power-law works across all the phases and over a large range of densities. It is also more physically motivated by the idea that a fraction of turbulent kinetic energy is converted to magnetic energy and does not require details of the geometry of the gas packet and local magnetic field orientation as usually necessary before applying the $B$ -- $n$ relations. Furthermore, using the probability distribution function of density, magnetic fields, and turbulent velocities (\Fig{fig:pdf}, \Tab{tab:stat}, and \Sec{sec:pdf}), we show that the magnetic field fluctuations are decided by both the density and turbulent velocity fluctuations. Finally, we also compute the typical plasma beta (\Tab{tab:beta} and \Sec{sec:beta}) and show that the magnetic pressure is comparable to the thermal pressure in all the ISM phases. In conclusion, our work demonstrates that magnetic fields are a dynamically important component of the ionised, atomic, and molecular ISM, determined by both gas density and turbulent velocity.

\section*{Acknowledgements}
\rev{We thank the anonymous referee for their fast and productive report.} A.~S.~thanks Christoph Federrath, Bryan M. Gaensler, Yik Ki Ma,  Antoine Marchal, \rev{Aris Tritsis}, Richard M. Crutcher, Enrique V\'{a}zquez-Semadeni, Timothy Robishaw, Alex S. Hill, and \rev{Chong-Chong He} for useful discussions. We thank Kate Pattle for providing the data used in \App{sec:emagekindcf}. The authors acknowledge Interstellar Institute's program ``II6'' and the Paris-Saclay University's Institut Pascal for hosting discussions that nourished the development of the ideas behind this work. \rev{A.~S.~acknowledges support from the Australian Research Council's Discovery Early Career Researcher Award (DECRA, project~DE250100003).} We also acknowledge funding provided by the Australian Research Council (Discovery Project  DP220101558 and Laureate Fellowship FL210100039 awarded to N.~M.~Mc-G.). 

\section*{Data Availability}
This work uses available data from the literature: Zeeman measurements from \citet{CrutcherEA2010}, pulsars data from \href{https://www.atnf.csiro.au/research/pulsar/psrcat}{https://www.atnf.csiro.au/research/pulsar/psrcat} \citep[version 2.4.0,][]{ManchesterEA2005}, H$\alpha$ data from \href{https://www.astro.wisc.edu/research/research-areas/galactic-astronomy/wham/wham-sky-survey}{https://www.astro.wisc.edu/research/research-areas/galactic-astronomy/wham/wham-sky-survey} \citep{HaffnerEA2003, HaffnerEA2010}, HI data from \href{https://www.astro.uni-bonn.de/hisurvey/AllSky_gauss}{https://www.astro.uni-bonn.de/hisurvey/AllSky\_gauss} \citep{McClure-GriffithsEA2009, KalberlaEA2010, WinkelEA2016}, and Davis-Chandrasekhar-Fermi (DCF) measurements (used in \App{sec:emagekindcf}) from \citet{PattleEA2023}. \rev{The multiphase ISM simulation data used in \Sec{sec:sim} is taken from \citet{SetaF2022} and \citet{Kado-FongEA2020}.} The analysed data underlying this article will be shared on a reasonable request to the corresponding author, Amit Seta (\href{mailto:amit.seta@anu.adu.au}{amit.seta@anu.adu.au}). 

\bibliographystyle{mnras}
\bibliography{magmpism} 


\appendix

\rev{

\section{Representative H$\alpha$ and HI spectra} \label{sec:spectra}
\begin{figure*}
\includegraphics[width=\columnwidth]{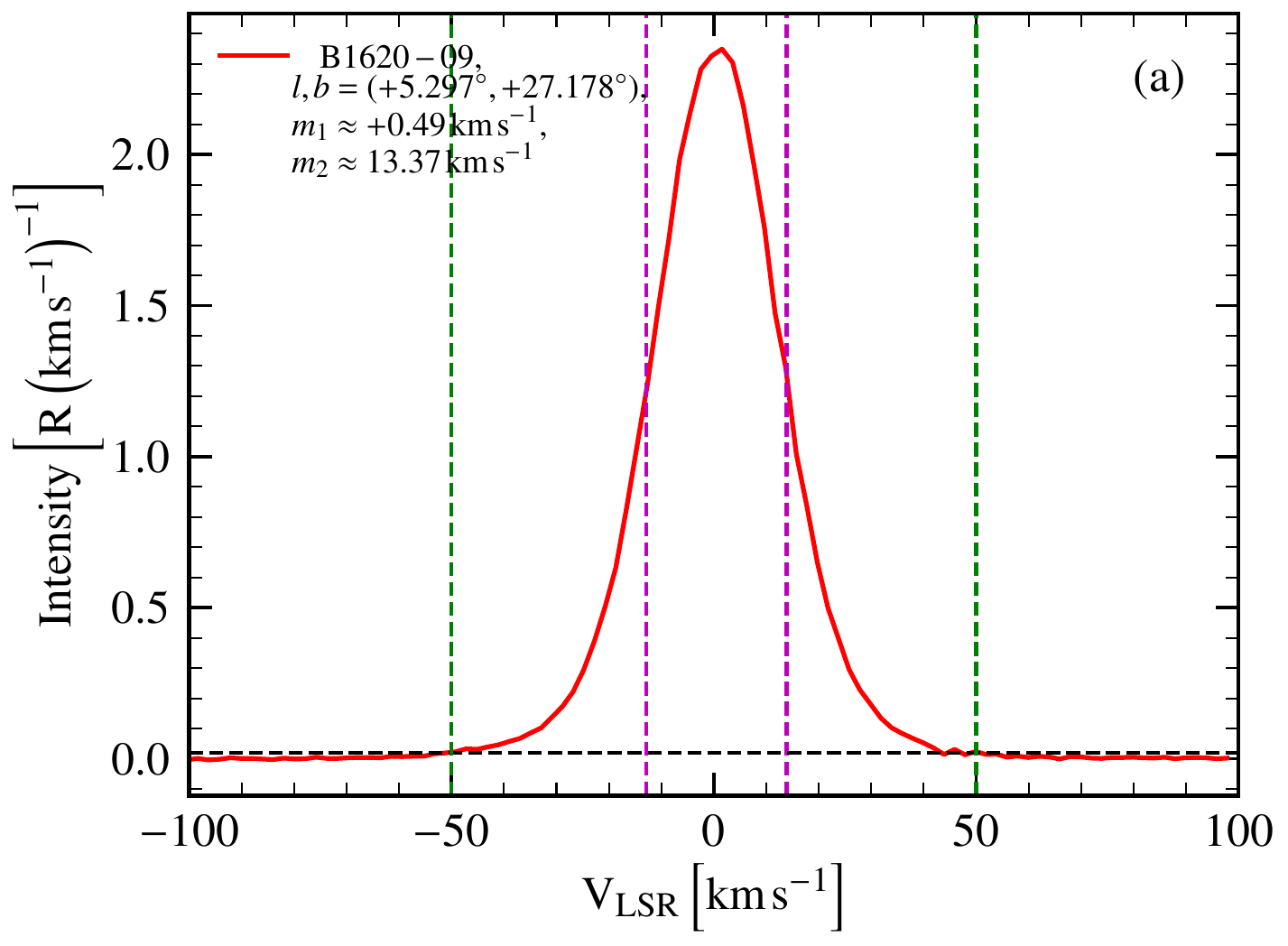}  \hspace{0.5cm}
\includegraphics[width=\columnwidth]{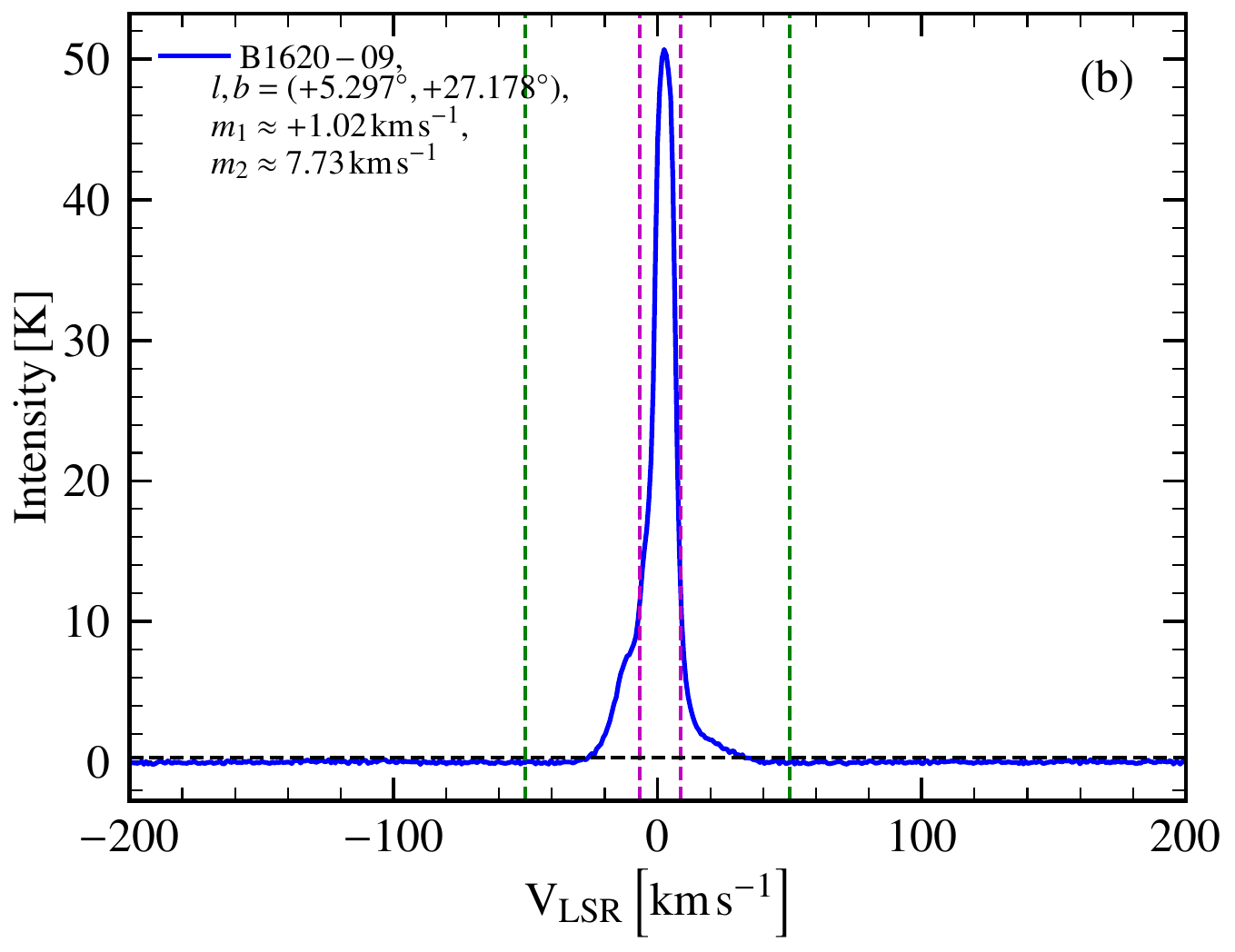} \\
\includegraphics[width=\columnwidth]{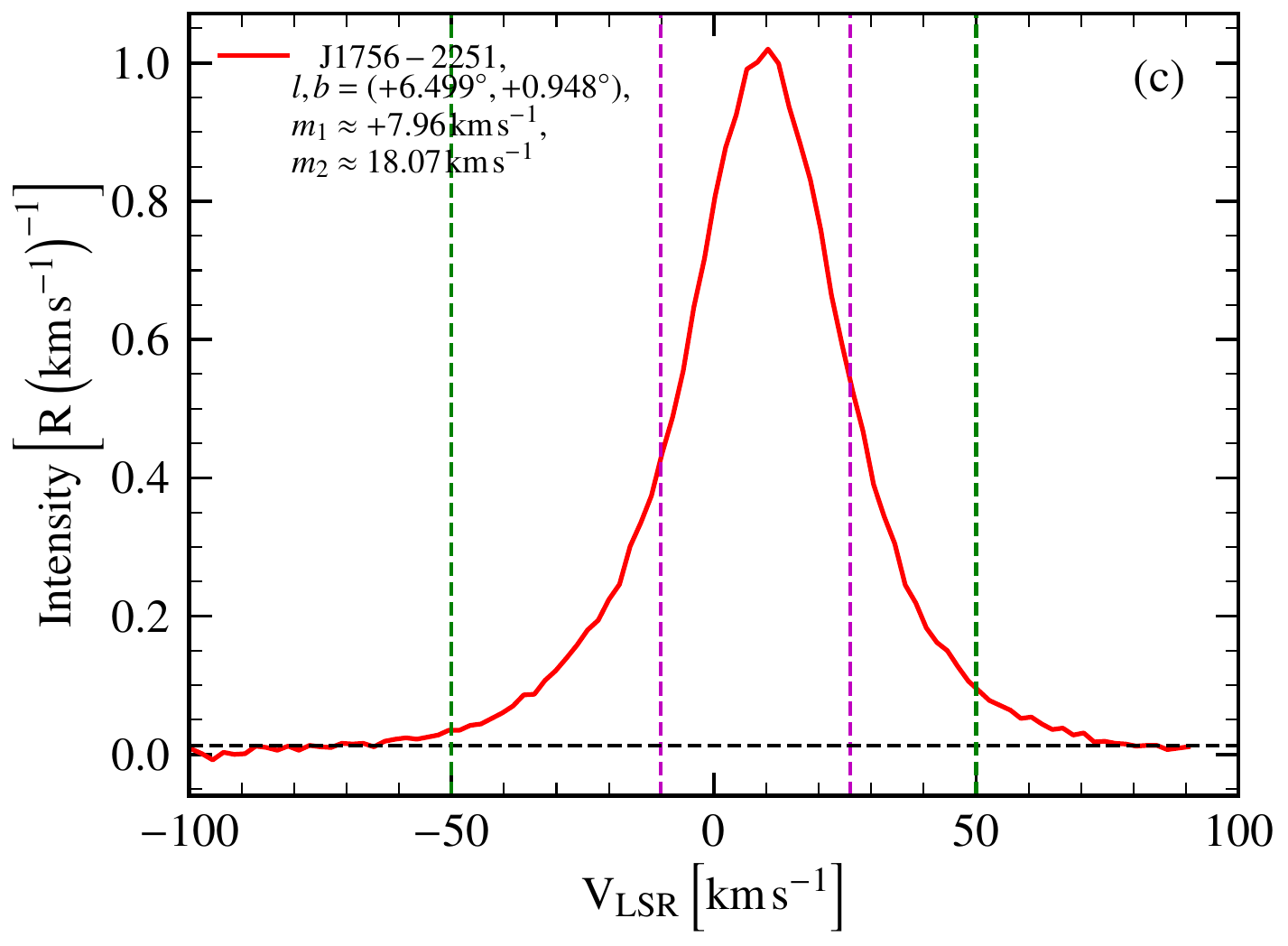}  \hspace{0.5cm}
\includegraphics[width=\columnwidth]{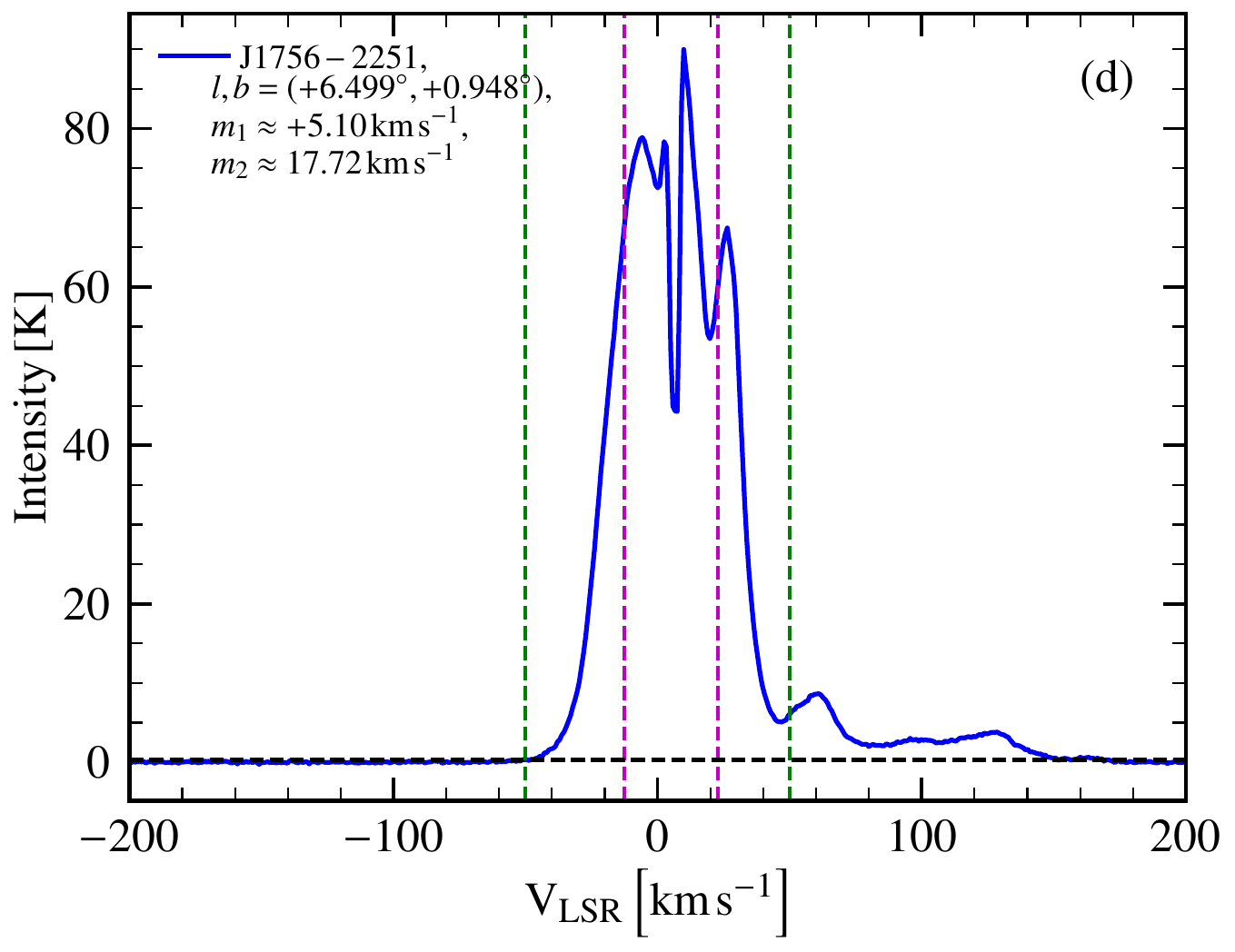}
\caption{Representative H$\alpha$ (a, c; red) and HI (b, d; blue) spectra for two pulsars: B1620 -- 09 (a, b; located away from the Galactic plane) and J1756 -- 2251 (c, d; closer to the Galactic plane). The vertical dashed green lines mark the range of Milky Way structures, i.e., the local standard of rest velocities within $\pm 50,\km,\s^{-1}$, over which the first ($m_{1}$) and second ($m_{2}$) moments are computed (values given in the legend). The horizontal dashed black line represents three times the noise level, above which the moments are computed (see \Sec{sec:vdata} for numerical values and further details). Finally, the vertical dashed magenta lines indicate the range of $m_{2}$, which is taken as the equivalent velocity width, $\dv$, along the pulsar line of sight, from which the turbulent velocity, $\dvt$, is derived. For both pulsars, the H$\alpha$ spectra exhibit a single component, possibly due to the large velocity channel width of $12\,\km\,\s^{-1}$. In contrast, the HI spectra (with a finer velocity channel width of $\sim 1\,\km\,\s^{-1}$) reveal more components and greater complexity, particularly for the pulsar located closer to the Galactic plane.}
\label{fig:spectra}
\end{figure*}

As described in \Sec{sec:vdata}, we use H$\alpha$ and HI spectra at pulsar locations to compute the corresponding second moment, $m_{2}$, using \Eq{eq:m2} within a local standard of rest velocity range of $\pm 50,\km,\s^{-1}$. This allows us to derive the equivalent velocity width and, ultimately, the turbulent velocity along the pulsar sightlines (further detailed in \Sec{sec:dvt}).

In \Fig{fig:spectra}, we present representative H$\alpha$ and HI spectra for two pulsars: B1620 -- 09 (located away from the Galactic plane) and J1756 -- 2251 (closer to the Galactic plane). In both cases, the H$\alpha$ spectra exhibit a single broad peak, likely due to the large velocity channel width of $12\,\km\,\s^{-1}$, which may limit sensitivity to small-scale components. In contrast, the HI spectra, with a finer velocity channel width of $\sim 1\,\km\,\s^{-1}$, reveal significant differences in spectral structure, showing more complex features for pulsars near the Galactic plane. Despite these structural variations, we do not differentiate based on spectral complexity. Instead, we consistently use $m_{2}$ to determine the velocity width and subsequently derive the turbulent velocity.

\section{$\Emag$ -- $\Ekin$ relationship for only the Zeeman measurements} \label{sec:Zeeman}

\begin{figure*}
\centering
\includegraphics[width=13.5cm, height=10cm]{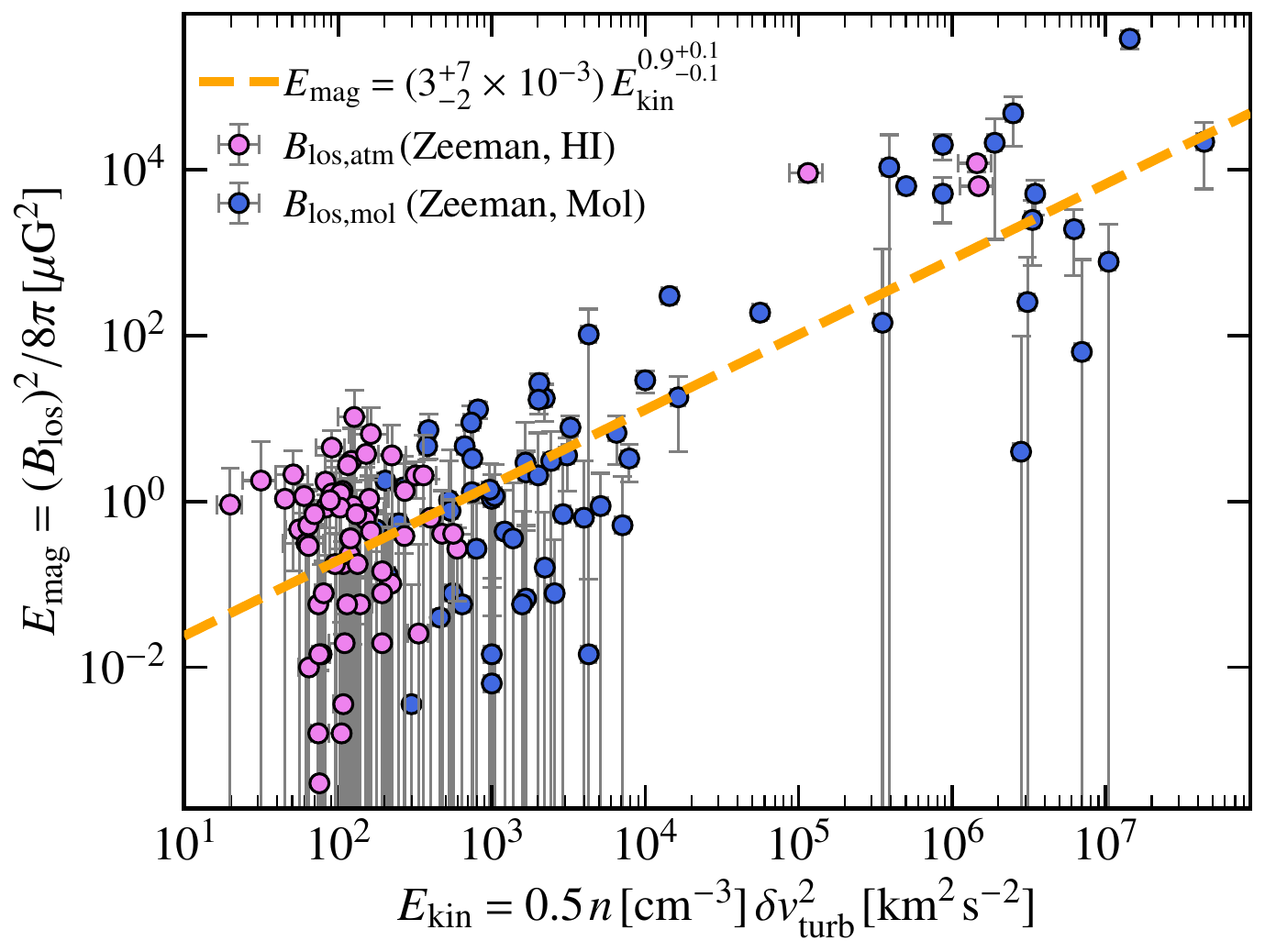}
\caption{The turbulent kinetic energy density, $\Ekin$, and magnetic energy density, $\Emag$, for the atomic (red, probed with $\ZHI$) and molecular (blue, probed with $\ZMol$) phases of the Milky Way’s ISM, along with a power-law fit (dashed orange line). The exponent of the fitted power law is closer to $1$ than in \Fig{fig:emagekin}, which includes all three phases (ionised, atomic, and molecular), where the analysis involving the ionised phase requires multiple assumptions and approximations (see \Sec{sec:dis}). This further supports the idea that $\Emag \propto \Ekin$ is a more fundamental physical relationship than $B$ -- $n$.}
\label{fig:Zeeman}
\end{figure*}

In the main text, we examined the $\Emag$ -- $\Ekin$ relation across all three phases of the Milky Way’s ISM--ionised, atomic, and molecular (\Fig{fig:emagekin}). Among these, the analysis involving the ionised ISM required the most assumptions and approximations (as discussed in \Sec{sec:dis}). In \Fig{fig:Zeeman}, we present the results of the power-law fit, $\Emag = \mathcal{C}_{E}\,\Ekin^{\alpha_{E}}$, derived exclusively from Zeeman measurements for the atomic and molecular phases. The fit suggests a scaling of $\Emag \propto \Ekin$ with an exponent $\alpha_{E}$ closer to $1$ than the values obtained in \Fig{fig:emagekin}, which includes all three phases. This further reinforces the idea that $\Emag \propto \Ekin$ represents a more fundamental physical relation than $B$ -- $n$.
}

\section{\rev{$\Emag$ -- $\Ekin$ relationship} including magnetic field measurements obtained using the Davis-Chandrasekhar-Fermi (DCF) method} \label{sec:emagekindcf}

\begin{figure*}
\includegraphics[width=\columnwidth]{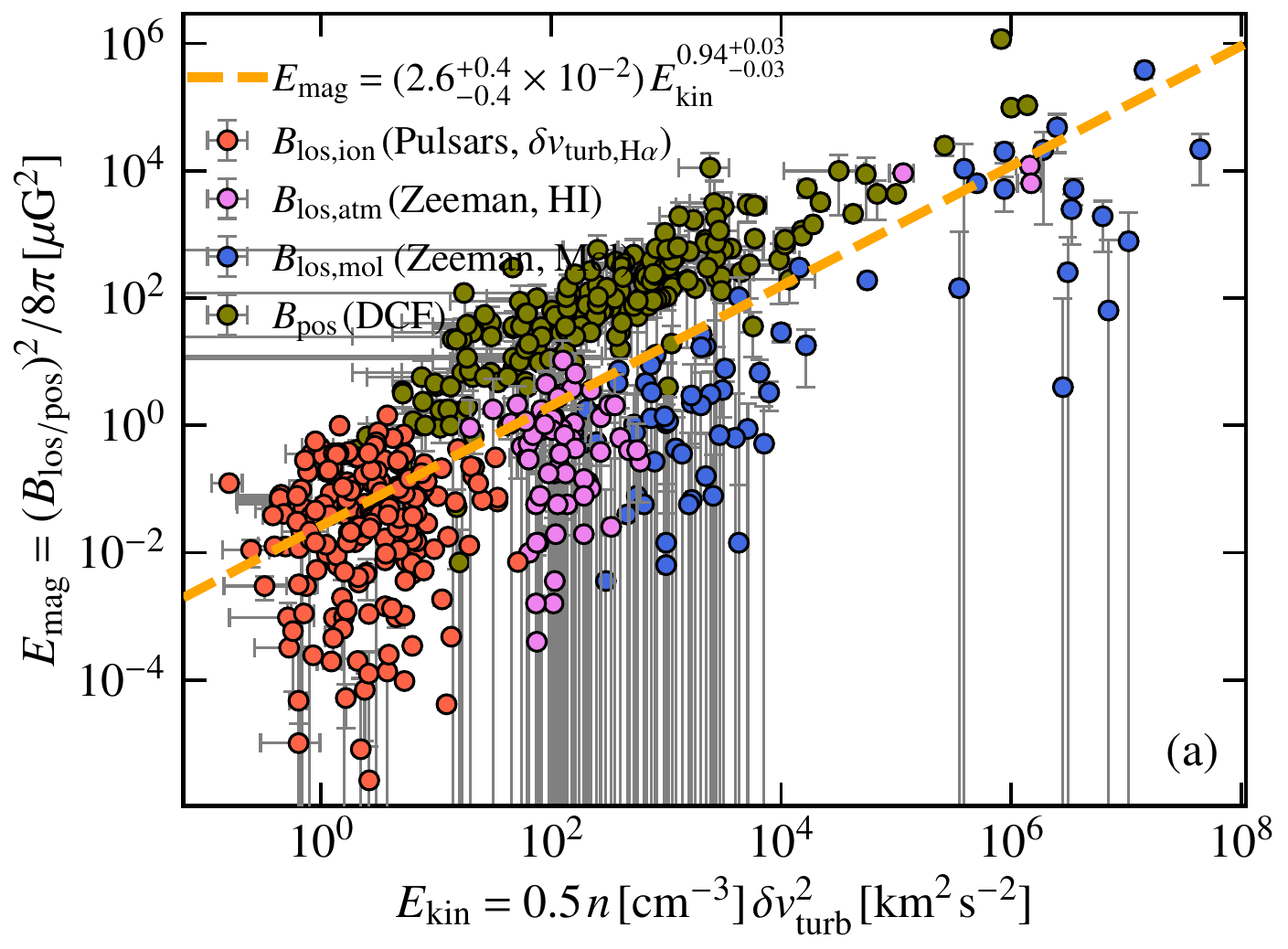}  \hspace{0.5cm}
\includegraphics[width=\columnwidth]{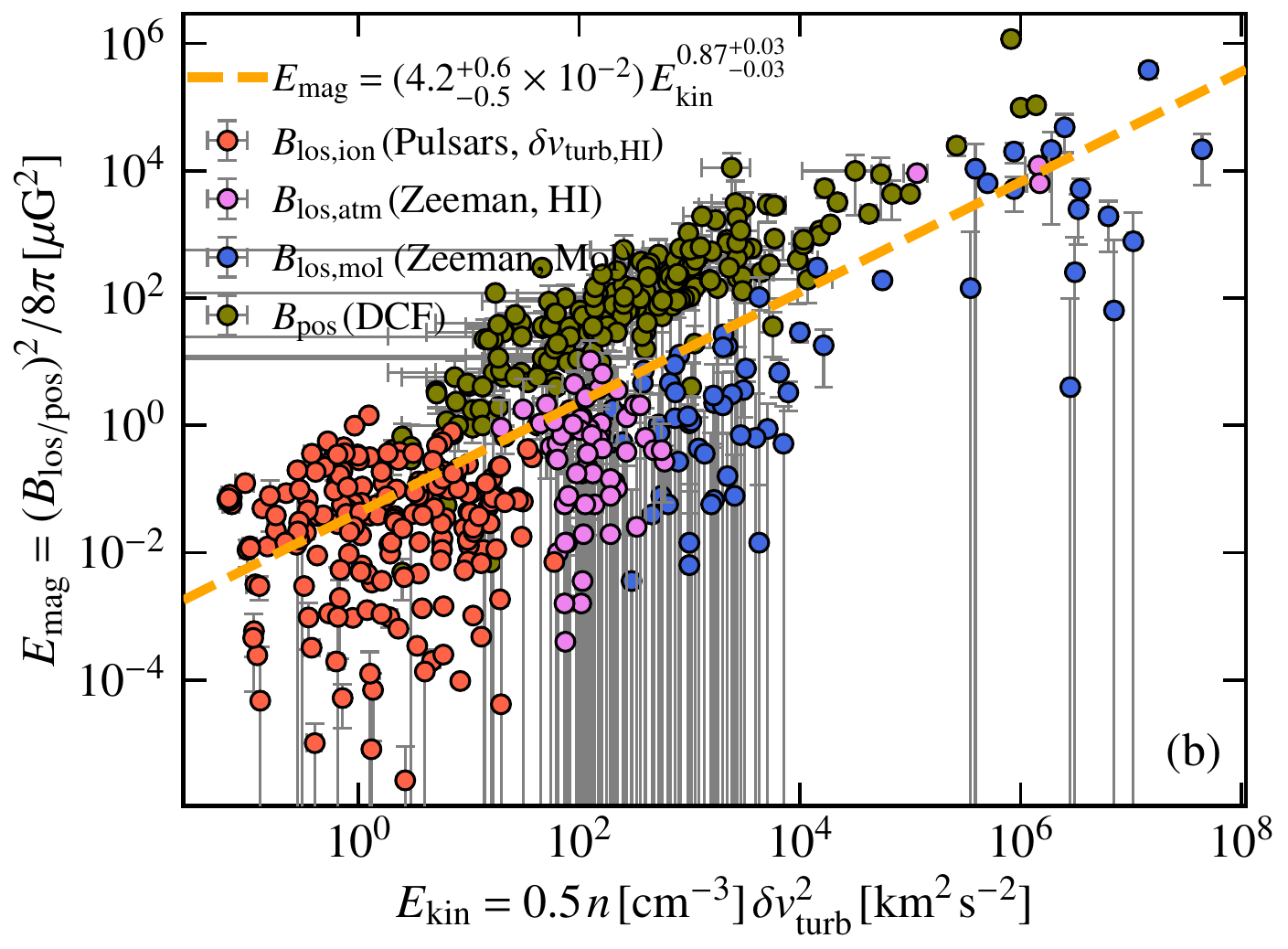}
\caption{Same as \Fig{fig:emagekin} but now also including the plane-of-sky magnetic field measurements, $\Bpos$, obtained using the Davis-Chandrasekhar-Fermi (DCF) method, data taken from \citealt{PattleEA2023}. The exponent of the power-law fit for both cases is closer to 1 in comparision to those in  \Fig{fig:emagekin}, further emphasising the $\Emag$ -- $\Ekin$ relation. However, the DCF method intrinsically assumes $\Emag = \Ekin$ to derive $\Bpos$ and thus this is expected.}
\label{fig:emagekindcf}
\end{figure*}

The Davis-Chandrasekhar-Fermi (DCF) method involves using dispersion in dust polarisation angle to estimate the plane-of-sky magnetic field, $\Bpos$ \citep{Davis1951, ChandrasekharF1953}. There are a few concerns about the applicability of this method, and some modifications of the classical method are also suggested \citep[see][for further details]{PattleF2019, SkalidisT2021, PattleEA2023}. We take the dust polarisation data ($n$ and $\dv$) with $\Bpos$ obtained using the DCF method from \citet{PattleEA2023} ($\nsamp=199$, only those measurements are chosen for which the uncertainties in $\Bpos$ and $\dv$ are available). Then, assuming the Mach number for these measurements, $\Mach_{\rm DCF}\approx1$, we compute $\dvt$ from $\dv$ using \Eq{eq:dvt}. Finally, $\Emag = B^{2}_{\rm pos} / 8 \pi$ and $\Ekin=0.5\,n\,\dvtt$ are computed. These are included with those obtained with the pulsars observations and Zeeman measurements and shown in \Fig{fig:emagekindcf}. The entire dataset is fitted with a power-law function. The exponents of the fit are significantly closer to 1 in comparision to those in \Fig{fig:emagekin}, which excludes the DCF values. However, this is expected, as inherently, the DCF method assumes $\Emag=\Ekin$ to derive $\Bpos$. Thus, this result does not convey significantly new information.


\bsp	
\label{lastpage}
\end{document}